\documentclass[12pt,prd,nofootinbib]{revtex4-1}
\usepackage[T1]{fontenc}
\usepackage{graphicx}
\usepackage{mathrsfs}
\usepackage{amssymb,amsmath}
\newcommand{\beq}[2]{\begin{equation}#1\label{#2}\end{equation}}
\newcommand{\ceq}[1]{(\ref{#1})}

\newfont{\mbld}{cmbx10 scaled 800}
\newfont{\cab}{cmsy10 scaled 1200}
\newfont{\scab}{cmsy10 scaled 1000}
\newfont{\bcall}{cmbsy10 scaled 1200}

\begin{document}
\title{Polymer plats and multicomponent anyon gases}
\author{Franco Ferrari$^{1}$, Jaros{\l}aw Paturej$^{1,2}$, Marcin
  Pi\c{a}tek$^{1,3}$ and Yani Zhao$^{1}$} 
\affiliation{ $^1$ CASA* and Institute of Physics, University of
  Szczecin, 
Wielkopolska 15, 70451 Szczecin, Poland
\\$^2$
Department of Chemistry, University of North Carolina, Chapel Hill,
North Carolina 27599-3290, United States 
\\
$^3$Bogoliubov Laboratory of Theoretical Physics, Joint Institute
for Nuclear Research, 141980, Dubna, Russia}
\begin{abstract}
Anyon systems are studied in connection with several interesting
applications including high $T_C$ superconductivity and topological
quantum computing. In this work we show that these systems can be
realized starting from directed polymers braided together
to form a nontrivial
link configuration belonging to the topological class of plats.
The statistical sum of a such plat is related here to the partition
function of a two-component anyon gas. The constraints 
that preserve the topological configuration of the plat are imposed
on the polymer trajectories using the so-called Gauss linking
number, a topological invariant that has already been well studied in
polymer physics. Due to these constraints, short-range
forces act on the monomers or, equivalently, on the anyon
quasiparticles in a way that closely resembles the appearance of
reaction forces in the constrained systems of classical mechanics.
If the polymers are homogeneous, the anyon system reaches a self-dual
point, in which these forces vanish exactly. A class of self-dual solutions
that minimize the energy of the anyons is derived. The
two anyon gas discussed here obeys an abelian
statistics, while for quantum computing it is known that nonabelian anyons are
necessary. However, this is a limitation due to the use of the Gauss
linking invariant to impose the topological constraints, which is a
poor topological invariant and is thus 
unable to capture the nonabelian characteristics of the braided
polymer chains.
A more refined treatment of the
topological constraints would require more sophisticated topological
invariants, but so far their application to the 
statistical mechanics of linked polymers
 is an open problem. 
\end{abstract}
\maketitle
\section{Introduction}
Knots and links are a fascinating subject and
are researched in connection with many concrete applications  both in
physics and biology \cite{grosberg,dna,katritch96,katritch97,krasnow,
  laurie,cieplak,marko,liu,wasserman,sumners,vologodski,orlandini,levene,
kurt,mehran,pp2,yan,arsuaga,arsuaga2,
metzler,pieranski,diao,sumners2,faddeev}.
In this paper we study the statistical mechanics of a system of two
entangled polymer rings. Mathematically, two or more entangled
polymers form what is called a
link. Single polymer rings form instead knots.
We will restrict ourselves to systems in the configurations of
$2s-$plats. Roughly speaking, $2s-$plats are knots or links obtained
by braiding together a set of $2s$ strings and connecting their ends
pairwise \cite{birman}. A physical realization of $2s-$plats could be that of two
rings topologically entangled together and with some of their points
attached to two membranes or surfaces located at different
heights. In nature $2s-$plats occur for example in the DNA of
living organisms \cite{diao,sumners,sumners2,darcy}. Indeed, it is believed
that most knots and links 
formed by DNA are in the class of $4-$plats \cite{sumners}. These
biological 
applications have inspired the research of 
Ref.~\cite{FFPLA2004}, in which $4-$plats have been studied with the
methods of statistical mechanics and field theory. In particular, in
\cite{FFPLA2004} it has been established an
analogy between polymeric $4-$plats and anyons, showing in this way
the tight relations between two component systems of quasiparticles
and the theory of 
knots and links.
After the publication of \cite{FFPLA2004}, 
interesting applications of analogous anyon systems to
topological quantum 
computing has been proposed \cite{dassarma,nayak,wilczek}.
These applications are corroborated by the results of 
experiments concerning  the detection of
anyons obeying a nonabelian statistics, see for example
\cite{goldman}. While these results have appeared in 2005 and are
still under debate
\cite{wilczek,anyonsexp2013}, other systems in which non-abelian
anyon statistics could be present
have been discussed \cite{gurarie,dolev}.
In the present case,
the topology of the original two-polymer link is distinguished by the
Gauss linking invariant, which can be obtrained from the amplitudes of
an abelian BF model \cite{thomblau}. This implies that
the statistics of the quasiparticles treated here is purely
abelian. However, also abelian anyons may be exploited for quantum
computations as it has been argued in Ref.~\cite{abeliananyons}.

Motivated by these recent advances, we study here the general
case of $2s-$plats formed by two polymer rings.
Among all knot and link
configurations, the class of
$2s-$plats is very special. For instance, it is possible to decompose the
trajectory of a $2s-$plat into a set of $2s$ open subtrajectories that
can be further
interpreted as the trajectories of $2s$ polymer chains directed along a
special direction. 
Without losing generality, we may suppose that this direction
coincides with the $z-$axis.
When the system of directed polymers is mapped into a field theory,
a model describing anyon quasiparticles is obtained. The $z$ coordinate
can be related to
"time", while the monomer densities of the $2s$
directed polymers become the quasiparticle densities of a multicomponent
anyon model. 
A remarkable feature of polymers in $2s-$plat
configurations is that 
they admit
self-dual solutions in which
the energy of the system is minimized \cite{FFPLA2004}. 
Here we show that these solutions can be explicitly constructed by
solving a sinh-Gordon equation.
The conformations corresponding 
to such  solutions should be particularly stable and thus
observable, at least in principle. 
With the present technologies \cite{membranes}, in fact, it is possible to realize
polymer $2s-$plats in the laboratory.

Another advantage of restricting ourselves
to $2s-$plats configurations is that it is possible to distinguish
their topological states in a  more efficient way than
what one could achieve in the general case of two linked polymer rings.
Let us recall at this point that
the trajectories of real polymers are impenetrable and thus, 
if no rupture occurs, they are
bound to stay in the initial topological state while subjected
to thermal fluctuations. However, 
in the Edwards' model used here polymers are "phantom" \cite{edwards}. Without any
control, their trajectories are allowed to cross themselves and thus
the global topological configuration of the system
may change.
To find a powerful and reliable method 
in order to forbid such changes of topology
is the most difficult problem of the statistical mechanics of
polymer knots and links.
Up to now there is no analytical model that is able to deal with the
statistical mechanics of polymer knots.
For this reason, in  this work we assume that each polymer ring
composing the link can be 
in any knot configuration.
Only the topological configurations
of the link formed by polymers
 belonging to the $2s-$plat
will be distinguished. 
This goal is achieved by using
the Gauss linking number in order to impose the necessary topological
constraints. 
The Gauss linking number is a topological invariant
given in the form of a double contour integral, where the
contours coincide with the polymer trajectories.
Unfortunately, it
is  a weak topological invariant, so that 
many nonequivalent topological configurations 
characterized by the same value of the Gauss linking number are
allowed. However, 
once we restrict ourselves to a given $2s-$plat,
 we are implicitly imposing a much more stringent 
topological condition on the system. Indeed, its topological states
are in this way not only limited by the value of the Gauss linking
number, but are 
 also forced to vary within the much smaller set of states 
that are compatible with the structure of the $2s-$plat.

Even if the Gauss linking number is one of the simplest topological
invariants, its expression is very complicated.
As a consequence, after imposing the
topological constraints, the action of a system of
topologically entangled polymers becomes both nonlocal and
nonpolynomial. 
The nonlocality is due to the double contour 
integral over the polymer trajectories. The nonpolynomiality
arises from the fact that the integrand is a nonpolynomial
function of the components of the radius vectors determining the 
positions of the monomers in the space. 
The situation is somewhat reminiscent to that
of holomic constraints in the classical mechanics of particles. When
these constraints are
fixed by means of Lagrange multipliers,
within the particle action  new terms appear which are
related to the reaction forces.
The striking difference is that
topological constraints are  not holonomic and 
have "memory" in order to keep
track of the global conformation of the chain. This last property causes
the nonlocality of the action after fixing the constraints.
The price to be paid to recover locality 
and to have a standard action
is to introduce topological
fields which interact with the monomers in such a way that the
topological configuration of the link is preserved. To some extent,
these interactions may be considered as the equivalents of the
reaction forces in classical mechanics. The passage to the topological
field theory description
is not straightforward.
In particular, it requires to find
a topological field theory with an
amplitude of metric independent and gauge invariant operators from
which it is possible to 
isolate the particular topological invariant used to fix the
topological constraints.
If the 
topological invariant is the Gauss linking number and the
two topologically linked rings
are represented as continuous
curves embedded in the space and parametrized by their arc-lengths,
this task has been achieved in \cite{FELAPLB1998, FELAJPA1999}.
In the present case, the two rings are constructed out of
a set of $2s$ open subtrajectories 
parametrized by the $z$ coordinate and
not by the arc-length. This parametrization is very peculiar because
it identifies the 
parameter specifying the positions of the monomers with
one coordinate of the space in which the monomers are
fluctuating. For all the above reasons,
the passage to topological field theories
explained in \cite{FELAPLB1998, FELAJPA1999} cannot be
straightforwardly applied to the present situation and has required
a separate derivation.
As a result of this derivation, we have been able to show that the 
path integral expressing the probability function of a
$2s-$plat formed by two 
polymer rings entangled together
is equivalent to the correlation function
of a gas of $2s_1$ particles of type 1
and $2s_2$ particles of type 2, where $s_1+s_2=s$.
The interactions between these particles are
mediated by the vector fields of
an abelian BF model. This is a topological gauge field theory that has been
discussed in \cite{thomblau,blauthompson}. 
The particles 
are also subjected to short-range interactions whose origin is the
following.
The $2s$
directed paths composing the two-polymer link are treated here like
paths of directed polymers in random media 
\cite{Kardar,kamien}, which are subjected to quenched random potentials.
After integrating over the random noise according to the prescriptions of
Ref.~\cite{Kardar}, in the polymer action of the $2s$
directed polymers appear
potentials describing short-range forces acting on the monomers. 

The final passage to field theory is performed using 
 the analog in statistical mechanics of the second quantization
process.
To this purpose, we generalized the method used by de Gennes
and coworkers \cite{degennes} to achieve the field theory formulation
in the case of 
a single polymer chain subjected to short-range interactions to the
case of a set of $2s$ different polymers.
In the ``second quantized'' version of the statistical mechanics of the
$2s-$plat, the scalar fields  create and destroy monomers in different
positions of the space. The square module of such fields may be
related to the monomer density at a certain point. The BF fields
take into account the interactions necessary to keep the system in its
initial topological configuration. 
The abelian BF model has been quantized  in
the Coulomb gauge, because in this gauge the analogy with anyon field
theories becomes particularly explicit.
The obtained field theory is  a multicomponent model of
anyons
such those described for instance in \cite{wilczek2}.
This kind of theories exhibits the phenomenon of superconductivity. 
The
only difference in our case is that the scalar fields containing the
creation and 
annihilation operators for particles of type 1 and 2 are organized in 
replica multiplets, where at the end the limit of zero replicas should
be taken.

This paper is organized as follows.
First of all, to map the partition function of two linked polymer
rings into that 
of anyons, it is necessary to split their trajectories into a set of $2s$
subtrajectories, which in the anyon model describe  the evolution in
time of the
quasiparticles. 
The splitting procedure and the definition of a time variable
that is able to parametrize the $2s$ subtrajectories
is carefully described in Section~\ref{sectionII}.
A proof that it is possible to isolate from the amplitudes
of the
BF model the Gauss linking number 
also after splitting the trajectories and changing their parametrization,
is presented in Section~\ref{sectionIII}. The fact that after
quantizing the BF model in the Coulomb gauge it is still possible to
recover the Gauss linking number from the amplitudes of the holonomies
is shown in the particular case of a $4-$plat in Appendix~B. In
Section~\ref{sectionIV} 
the partition function of two linked polymers subjected to topological
constraints imposed with the help of the Gauss linking invariant is
transformed into a theory of $2s$ directed polymers interacting with
the magnetic-like fields of the BF model.
Contrarily to Ref.~\cite{FFPLA2004}, we
treat the $2s$ subtrajectories  as 
trajectories of real directed polymers. This requires the introduction of
random potentials which  complicates somewhat the passage to the anyon field
theory performed in Section~\ref{sectionV}. In the anyon formulation,
the 
densities of monomers associated to the two original
polymer rings can be regarded as the densities of anyon quasiparticles
of type 1 and 2 interacting together. Thanks to a Bogomol'nyi
transformation, the interactions may be split into a self-dual part
and a part containing only short-range interactions.
Remarkably, the latter interactions  persists even if  the
short-range interactions coming from the random media are switched
off.
This is an effect of the
presence of the topological constraint.
In Section~\ref{sectionVI} it is reviewed for completeness
the case of a $4-$plat studied in \cite{FFPLA2004}.  The static
configurations of the anyon densities that minimize the
Hamiltonian are computed.
It is shown that the anyon model admits static self-dual points.
The nature of the density configurations corresponding to these
self-dual points is analyzed in  Section~\ref{sectionVII}.
We prove that the solutions of the classical equations of motion that
minimize the static Hamiltonian are self-dual configurations, whose
exact form can be obtained after solving a sinh-Gordon
equation. Finally, our conclusions are drawn in
Section~\ref{sectionVIII}. 
\newpage
\section{Polymers as $2s-$plats}\label{sectionII}
Let's consider two closed loops $\Gamma_1$ and $\Gamma_2$ of lengths
$L_1$ and $L_2$ respectively in a three dimensional space with
coordinates $(\mathbf r,z)$. 
The vectors $\mathbf r=(x,y)$ span the two
dimensional space $\mathbb {R}^2$.
$z$ will play  later on
the role of  time. The two loops will be labeled by
using a indices the first letters of the latin alphabet: $a,b,\ldots=1,2$.
We will assume  that $\Gamma_1$ and $\Gamma_2$ form a $2s-$plat. For
convenience, we briefly review what is a $2s-$plat. First of all, we
recall that a single closed trajectory is from the mathematical point
of view a knot, while a system of knots linked together forms
a link. 
\begin{figure}
\centering
\begin{minipage}{6cm}
\includegraphics[width=5.9cm]{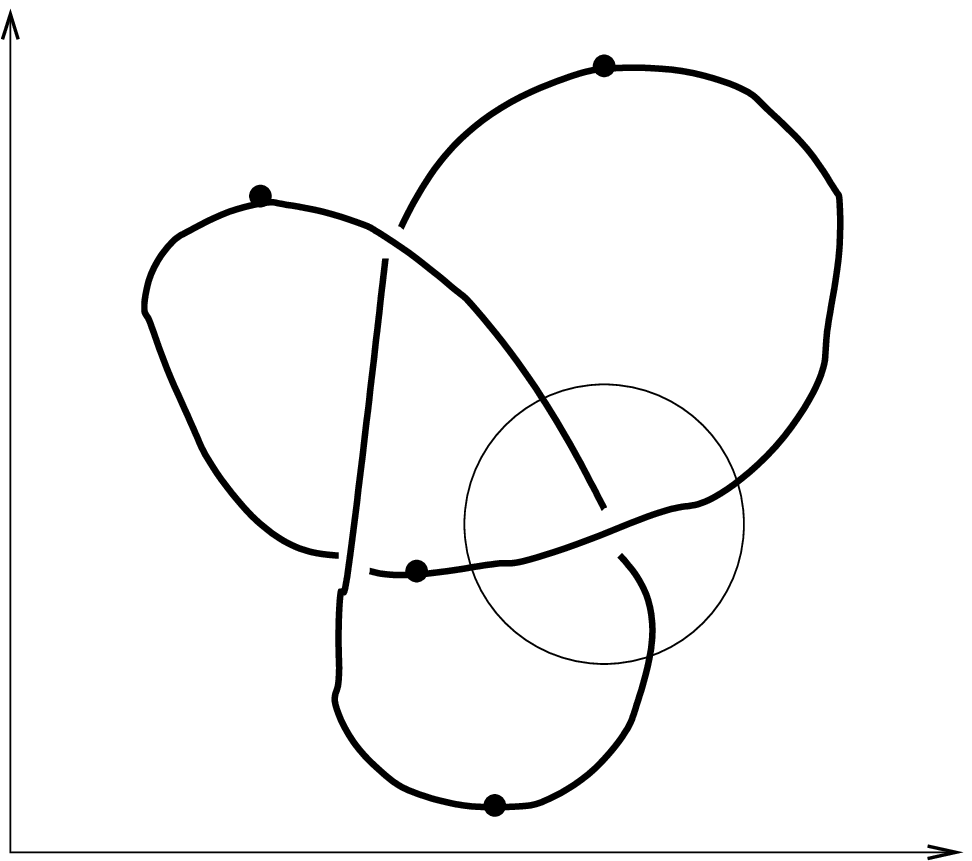}
\caption{Representation of a treefoil knot in terms as a
  two-dimensional
diagram.}\label{treefoil}
\end{minipage}\hspace{1cm}
\begin{minipage}{6cm}
\vspace*{2.6cm}
\includegraphics[width=5.9cm]{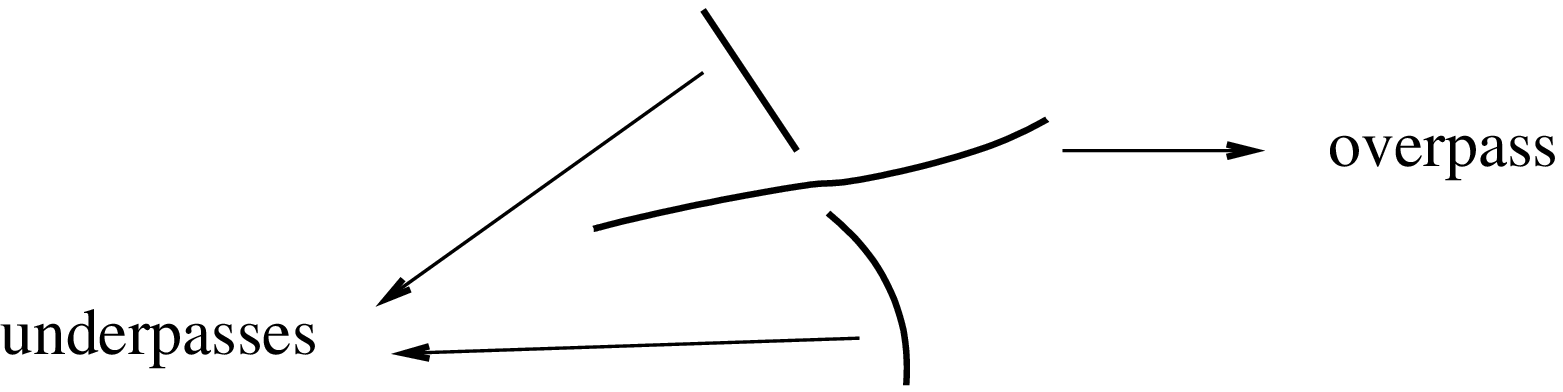}
\vspace{2cm}
\caption{The figure shows one of the crossings which are present
in the diagram of the treefoil knot of Fig.~\ref{treefoil}}\label{overunder}
\end{minipage}
\end{figure}
Knots and links may be represented after a projection onto a
plane by diagrams like those of Fig.~\ref{treefoil} and \ref{basiclink},
in which the 
original three-dimensional structure is 
simulated by a system of crossings, see Fig.~\ref{overunder}. 
\begin{figure}[bhpt]
\centering
\includegraphics[scale=.33]{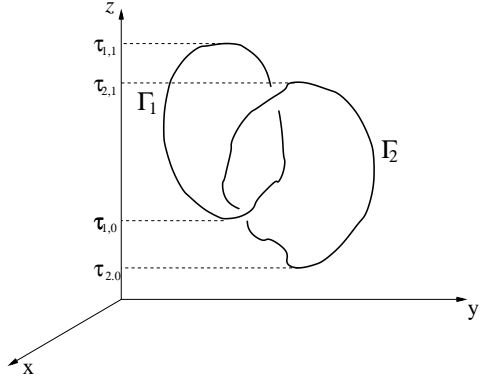}\\
\caption{A link formed by two polymers $P_1$ and $P_2$.}\label{basiclink}
\end{figure}
Each crossing is composed by
three arcs, one 
overpass and two underpasses.
One may also realize that the treefoil diagram in Fig.~\ref{treefoil} is
characterized by two minima  and two
maxima. Two dimensional diagrams of this kind, deformed in such a way
that the number $2s$ of minima and maxima is the smallest possible and
the maxima and minima are aligned at the same heights $z_{Max}$
and $z_{Min}$ respectively, are called in knot theory
$2s-$plats~\footnote{Actually, to be rigorous one 
  should still require that 
  neither maxima nor minima occur at the crossing points.}.
The height of a $2s-$plat is measured here with respect
to the $z$ axis.
$2s-$plats are used to classify knots and links by dividing them into
classes characterized by the same value of $s$. The concept of
$2s-$plats arises naturally in biochemistry, see
e.~g.~\cite{sumners}. In the present case, with some abuse of language,
we will call $2s-$plats also the two dimensional diagrams in which
maxima and minima are not 
aligned, like for instance in Fig.~\ref{basiclink}. 
Let us denote
with the symbols $ \tau_{a,I_a}$, $I_a=0,\ldots, 2s_a-1$,
the heights of the maxima and minima of each
trajectory $\Gamma_a$, for $a=1,2$, 
Arbitrarily, we
choose $\tau_{1,0}$ and $\tau_{2,0}$ to be the heights
of the lowest minima on the trajectories $\Gamma_1$ and $\Gamma_2$
respectively. 
Starting from $\tau_{a,0}$, we select the orientation of
$\Gamma_a$ in such a way that, proceeding along the trajectory
according to that orientation, we will encounter in the order the
points $\tau_{a,1},\tau_{a,2},\ldots,\tau_{a,2s_a}$. Clearly,
$\tau_{a,1}$ is a point of maximum, $\tau_{a,2}$ one of minimum and so
on.
Moreover, we should put for consistency:
\begin{equation}
\tau_{a,2s_a}\equiv\tau_{a,0}
\end{equation}
The introduction of this double notation for the same height 
$\tau_{a,0}$ will be useful in the following in order to write
formulas in a more compact form.
In the following the $2s-$plats $\Gamma_1$ and $\Gamma_2$ will be
decomposed into
a set
of directed trajectories
$\Gamma_{a,I_a}$, $a=1,2$ and $I_a=0,\ldots, 2s_a-1$,
whose ends are made to coincide in such a
way that they form the topological configuration of two linked rings.
An example when $s=3$ is presented in Fig.~\ref{sectioning}.
\begin{figure}[bpht]
\centering
\includegraphics[scale=.5]{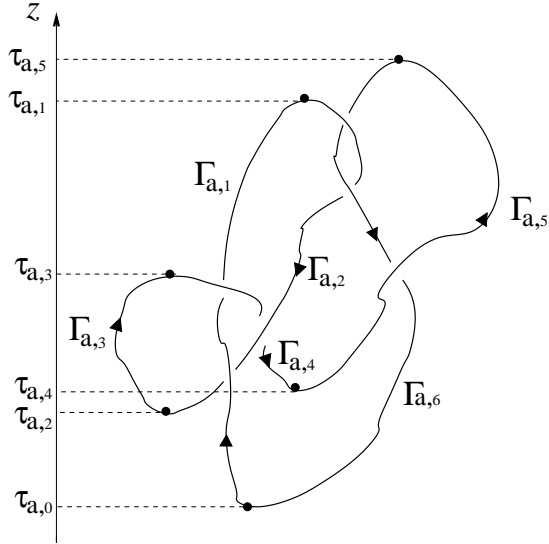}\\
\caption{Sectioning procedure for a $2s$-plat $\Gamma_a$ with
$s=3$.}\label{sectioning}
\end{figure}
In the general case the set of points belonging to $\Gamma_{a,I_a}$
can be described by the formula:
\begin{equation}
\Gamma_{a,I_a}=\left\{
\mathbf r_{a,I_a}(z_{a,I_a})\left|\begin{array}{c}
a=1,2;\qquad I_a=1,\ldots,2s_a\\
\left\{\begin{array}{rcl}
\tau_{a,I_a-1}\le z_{a,I_a}\le\tau_{a,I_a}\qquad
I_a\mbox{ odd}\\
\tau_{a,I_a}\le z_{a,I_a}\le\tau_{a,I_a-1}\qquad
I_a\mbox{ even}
\end{array}
\right.
\end{array}
\right.\right\}\label{bcondone}
\end{equation}
where the additional conditions:
\begin{eqnarray}
\mathbf r_{a,I_a}(\tau_{a,I_a})&=&\mathbf
r_{a,I_a+1}(\tau_{a,I_a})\qquad I_a=1,\ldots,2s_a-1\label{bcondtwoa}\\
\mathbf r_{a,1}(\tau_{a,0})&=&\mathbf r_{a,2s_a}(\tau_{a,0})
\label{bcondtwo}
\end{eqnarray}
which connect together the subtrajectories $\Gamma_{a,I_a}$ so that 
the loop $\Gamma_a$ is reconstructed, are understood.
In Eq.~(\ref{bcondone}) $\mathbf r_{a,I_a}(z_{a,I_a})$ represents the
projection of the trajectory $\Gamma_{a,I_a}$ onto the plane $x,y$
transverse to the longitudinal direction $z_{a,I_a}$.
Let us note that we are using the same indexes $I_a$ to label the
trajectories $\Gamma_{a,I_a}$ and the points $
\tau_{a,I_a}$. However, in the first case $I_a=1,\ldots,2s_a$, while
in the second case we have chosen $I_a=0,\ldots,2s_a-1$.
In the case of the variables $z_{a,I_a}$'s, the range of $I_a$
is the same as that of the $\Gamma_{a,I_a}$'s.

We notice that the variables $z_{a,I_a}$'s are always
growing and  do not take automatically into account the fact that
the whole chain is continuous and has a given orientation. 
Better variables, both with respect to the continuity and orientation,
are the following $t_{a,I_a}$'s:
\begin{eqnarray}
t_{a,I_a}=z_{a,I_a}&\qquad&\mbox{when }I_a\mbox{ is
  odd}\label{transfone}\\ 
t_{a,I_a}=-(z_{a,I_a}-\tau_{a,I_a})
+ \tau_{a,I_a-1}
&\qquad&\mbox{when }I_a\mbox{ is even}\label{transftwo}
\end{eqnarray}
Assuming for instance that $I_a$ is odd, for two consecutive
trajectories $\Gamma_{a,I_a}$ and $\Gamma_{a,I_{a+1}}$, we have that:
\begin{equation}
\tau_{a,I_a-1}\le t_{a,I_a+}\le\tau_{a,I_a}
\end{equation}
and
\begin{equation}
\tau_{a,I_a}\ge t_{a,I_a+1}\ge\tau_{a,I_a+1}
\end{equation}
According to the above conventions, trajectories labeled by odd
$I_a$'s are oriented from a point of minimum to a point of maximum,
while trajectories with even values of $I_a$ go from a point of
maximum to a point of minimum.
In the new coordinate $t_{a,I_a}$, the trajectory $\Gamma_{a,I_a}$
becomes parametrized as follows:
\begin{equation}
\Gamma_{a,I_a}=\left\{
\mathbf r_{a,I_a}( t_{a,I_a})\left|\begin{array}{c}
a=1,2;\qquad I_a=1,\ldots,2s_a\\
\left\{\begin{array}{rcl}
\tau_{a,I_a-1}\le  t_{a,I_a}\le\tau_{a,I_a}\qquad
I_a\mbox{ odd}\\
\tau_{a,I_a-1}\ge  t_{a,I_a}\ge\tau_{a,I_a}\qquad
I_a\mbox{ even}
\end{array}
\right.
\end{array}
\right.\right\}\label{conddd}
\end{equation}
where the boundary conditions \ceq{bcondtwoa} and \ceq{bcondtwo} are
understood. 
\section{The Gauss linking number and the abelian BF field
  theory}\label{sectionIII}
To express the topological properties of the system of two linked
loops $\Gamma_1$ and 
$\Gamma_2$, we use as a topological invariant the Gaussian linking
number:
\begin{equation}
\chi(\Gamma_{1},\Gamma_{2})=\frac{1}{4\pi}\varepsilon_{\mu\nu\rho}
\oint_{\Gamma_{1}} d\tilde x^{\mu}_{1}(\sigma_1)\oint_{\Gamma_{2}}
 d\tilde x^{\nu}_{2}(\sigma_2)\frac{(\tilde x_{1}(\sigma_1)
-\tilde x_{2}(\sigma_2))^{\rho} 
}{|\tilde x_{1}(\sigma_1)-\tilde x_{2}(\sigma_2)|^{3}} \label{ln}
\end{equation}
where the $\tilde x^\mu_a(\sigma_a)$'s, $a=1,2$, are closed curves
representing the loops $\Gamma_1$ and $\Gamma_2$
 in the three dimensional space. The
variables $\sigma_1$ and $\sigma_2$ used to parametrize 
$\Gamma_1$ and $\Gamma_2$ represent the respective arc-lengths of the
two loops. They are defined in such a way that $0\le\sigma_a\le L_a$.
In the following, the trajectories of the two loops will be
topologically constrained
by the condition:
\begin{equation}
m=\chi(\Gamma_1,\Gamma_2)\label{topconst}
\end{equation}
$m$ being a given integer. The above constraints is imposed by
inserting
the Dirac delta function
$\delta(m=\chi(\Gamma_1,\Gamma_2))$ in
the partition function of the $2s-$plat, where
the statistical sum over all conformations of $\Gamma_1$ and
$\Gamma_2$ is performed. Of course, the analytical treatment of such a
delta function in a path integral is difficult. Some
simplification is obtained by 
passing to
the Fourier representation:
\begin{equation}
\delta(m-\chi(\Gamma_1,\Gamma_2))=\int_{-\infty}^{+\infty}\frac{d\lambda}{\sqrt{2\pi}}
e^{-i\lambda(m-\chi(\Gamma_1,\Gamma_2))}\label{deltaft}
\end{equation}
However, even in the Fourier representation,
the difficulty of having to deal with the Gauss linking number in the
exponent appearing in the right hand side of Eq.~(\ref{deltaft})
remains. Formally,
this  topological invariant introduces a term that  resembles
the potential of a two-body interaction. However, this
potential is both nonlocal and 
nonpolynomial. It is for that reason that the treatment of the Gauss
linking number in any microscopical model of topologically entangled
polymers is usually very 
complicated. The best strategy do deal with it so far consists
in rewriting the delta
function
$\delta(m-\chi(\Gamma_1,\Gamma_2))$ as a correlation function of the
holonomies of a local field theory, namely the so-called abelian
BF-model \cite{FELAPLB1998,FELAJPA1999,thompsonblau}:
\begin{equation}
\delta(m-\chi(\Gamma_1,\Gamma_2))=\int_{-\infty}^{+\infty}d\lambda
e^{-i\lambda m}{\tilde {\cal Z}}_{BF}(\lambda)\label{maintop}
\end{equation}
where
\begin{eqnarray}
{\tilde{\cal Z}_{BF}}(\lambda)&=&\int{\cal D}\tilde B_\mu(x){\cal
  D}\tilde C_\mu(x)
e^{-iS_{BF}[\tilde B,\tilde C]}\nonumber\\
&\times&e^{-i\tilde c_1\oint_{\Gamma_1}d\tilde x^\mu_1(\sigma_1)\tilde
  B_\mu(\tilde{x}_1(\sigma_1))}
e^{
-i\tilde c_2
\oint_{\Gamma_2}
d\tilde x^\mu_2
(\sigma_2)\tilde C_\mu(\tilde{x}_2(\sigma_2))
}\label{c2a3}
\end{eqnarray}
In the above equation we have put $x\equiv(\mathbf x,t)$ to be
dummy integration variables in the three dimensional space. Moreover,
$S_{BF}[\tilde B,\tilde C]$ denotes the action of the abelian BF-model:
\begin{equation}
S_{BF}[\tilde B,\tilde C]=\frac\kappa{4\pi}\int d^3x\tilde
B_\mu(x)\partial_\nu \tilde C_\rho(x)\epsilon^{\mu\nu\rho} \label{bfmodelaction}
\end{equation}
$\epsilon^{\mu\nu\rho} $, $\mu,\nu,\rho=1,2,3$, being the completely
antisymmetric 
$\epsilon-$tensor density defined by the condition $\epsilon^{123}=1$.
$\kappa$ is the coupling constant of the BF-model. 
Finally, 
the constants $\tilde c_1$ and
$\tilde c_2$  are given by:
\begin{equation}
\tilde c_1=\lambda\qquad\qquad \tilde c_2=\frac\kappa{8\pi^2}
\end{equation}
While there is some freedom in choosing $\tilde c_1$ and $\tilde c_2$,
one unavoidable
requirement in order that Eq.~(\ref{maintop}) will be satisfied is that one of
these parameters should be linearly dependent on $\kappa$.
In this way, it is easy to check that $\kappa$ may be completely
eliminated from Eq.~(\ref{c2a3}) by performing a
rescaling of one of
the two fields $\tilde B_\mu$ and $\tilde C_\mu$.
This is an expected result, because $\kappa$
did not appear in
the left hand 
side of Eq.~(\ref{maintop}), so that it
cannot be a new parameter of the theory.
By introducing the currents:
\begin{equation}
\tilde J_a^\mu( x)=\tilde
c_a\oint_{\Gamma_a}d\tilde x_a^\mu (\sigma_a)\delta^{(3)}(
x- \tilde{ x}_a(\sigma_a))\qquad a=1,2\label{tildecurrent}
\end{equation}
$\tilde {\cal Z}_{BF}(\lambda)$ may be rewritten in the more compact way:
\begin{eqnarray}
\tilde{\cal Z}_{BF}(\lambda)&=& \int{\cal D}\tilde B_\mu(x)
{\cal D}\tilde C_\mu(x)e^{-iS_{BF}[\tilde B,\tilde C]}
e^{-i\int d^3x\left[
\tilde J_1^\mu
(x)\tilde B_\mu(x)+
\tilde J_2^\mu
(x)\tilde C_\mu(x)
\right]}\label{maintopcurrents}
\end{eqnarray}

In all the above discussion, the two trajectories
$\Gamma_1$ and $\Gamma_2$ have been
parametrized with the help of the arc-lengths $\sigma_1$ and $\sigma_2$.
However,
in the case of a $2s-$plat, the loops $\Gamma_a$ are realized
as a set of open 
paths $\Gamma_{a,I_a}$ connected together by the conditions
(\ref{bcondtwoa}--\ref{bcondtwo}). 
The subtrajectories $\Gamma_{a,I_a}$'s are
directed paths
$\mathbf
r_{a,I_a}(t_{a,I_a})=(x^1_{a,I_a}(t_{a,I_a}),x^2_{a,I_a}(t_{a,I_a}))$
parametrized
by the new variables $t_{a,I_a}$, which are connected to the
third spatial coordinates
$z_{a,I_a}=x^3$ by the relations \ceq{transfone} and
  \ceq{transftwo}.
Due to this difference of parametrization, the above method to
express the Gauss linking number based on the BF-model, in particular
Eq.~(\ref{maintop}),
 should be
changed appropriately. Our starting point is the new partition
function:
\begin{eqnarray}
{\cal Z}_{BF}(\lambda)&=& \int{\cal D} B_\mu(x)
{\cal D} C_\mu(x)e^{-S_{BF}[ B, C]}
e^{-i\int d^3x\left[
J_1^\mu
(x) B_\mu(x)+
J_2^\mu
(x) C_\mu(x)
\right]}\label{zlambdawotilde}
\end{eqnarray}
where $S_{BF}[ B, C]$
coincides with the action (\ref{bfmodelaction}) but with the fields
$\tilde B,\tilde C$ renominated $B,C$ and
\begin{equation}
J_a^\mu(x)=\tilde
c_a\sum_{I_a=1}^{2s_a}\int_{\tau_{a,Ia-1}}^{\tau_{a,I_a}}dx_{a,I_a}^\mu(\tau_{a,I_a})
\delta^{(3)}(x-x_{a,I_a}(t_{a,I_a}))
\label{currentwotilde}
\end{equation}
Let us note that 
the  transformation
  $z_{a,I_a}\rightarrow t_{a,I_a}$ provided by Eqs.~\ceq{transfone} and
  \ceq{transftwo} leaves the BF action and the source
  terms unaffected, so that it does not change the form of the path
  integral $Z_{BF}(\lambda)$. For this reason, starting
from Eq.~(\ref{currentwotilde}), 
the directed paths $\Gamma_{a,I_a}$ are parametrized
with the variables $t_{a,I_a}$ instead of $z_{a,I_a}$.

To show that the partition function ${\cal Z}_{BF}(\lambda)$
of Eq.~(\ref{zlambdawotilde}) coincides with the partition function 
$\tilde{\cal Z}_{BF}(\lambda)$ of Eq.~(\ref{maintopcurrents}),
we introduce new variables $\sigma_{a,I_a}$ as
follows.
Let $\varsigma_{a,I_a}$, $I_a=0,\ldots,2s_a-1$ be the arc-length
on $\Gamma_a$ of the point of maximum or minimum located at the height
$ \tau_{a,I_a}$. The $\sigma_{a,I_a}$'s are defined in such a
way that they span the intervals:
\begin{equation}
\varsigma_{a,I_a}\le\sigma_{a,I_a}\le\varsigma_{a,I_a+1}
\end{equation}
As a brief digression, even if it is not necessary for the present discussion,
let us define the arc-length $\sigma'_{a,I_a}$ of each trajectory
$\Gamma_{a,I_a}$. It is easy to show that  $\sigma'_{a,I_a}$ is given by:
\begin{equation}
\sigma'_{a,I_a}=\sigma_{a,I_a}-\varsigma_{a,I_a}
\end{equation}
As a matter of fact, $\sigma'_{a,I_a}$ ranges in the interval
$[0,\varsigma_{a,I_a+1}-\varsigma_{a,I_a}]$ and
$\varsigma_{a,I_a+1}-\varsigma_{a,I_a}$ is the total length of 
$\Gamma_{a,I_a}$ for $a=1,2$ and $I_a=,\ldots,2s_a-1$.

On each subtrajectory $\Gamma_{a,I_a}$ defined by Eq.~(\ref{conddd}),
we can separately pass from the parameters $t_{a,I_a}$ to the
arc-length of $\Gamma_{a,I_a}$ by a transformation of the kind
$t_{a,I_a}=t_{a,I_a}(\sigma_{a,I_a})$. Putting
\begin{equation}
\tilde
x^\mu_{a,I_a}(\sigma_{a,Ia})=x^\mu_{a,I_a}(t_{a,I_a}(\sigma_{a,I_a}))
\end{equation}
we may rewrite the currents $J^\mu_a(x)$ as follows:
\begin{equation}
J_a^\mu(x)=\sum_{I_a=1}^{2s_a}{\tilde c}_a
\int_{\varsigma_{a,I_a-1}}^{\varsigma_{a,I_a}}d{\tilde x}_{a,I_a}^\mu(\sigma_{a,I_a})
\delta^{(3)}(
x-\tilde{x}_{a,I_a}(\sigma_{a,I_a}))\label{defghi}
\end{equation}
It is easy to realize that the sum over all values of $I_a$ in
Eq.~(\ref{defghi}) is 
equivalent to a contour integration over the whole loop
$\Gamma_a$. As a consequence:
\begin{equation}
J_a^{\mu}(x)=\tilde
c_a\oint_{\Gamma_a}d\tilde x_a^\mu(\sigma)\delta^{(3)}(x-
\tilde{x}_a(\sigma_a))\equiv\tilde J_a^\mu(x)
\end{equation}
i.~e. the currents
$J_a^\mu(x)$ coincide with the currents $\tilde
J_a^\mu(x)$ defined in
Eq.~(\ref{tildecurrent}). Therefore, in the partition function ${\cal 
  Z}_{BF}(\lambda)$ of Eq.~(\ref{zlambdawotilde})  the currents
$J_a^\mu(x)$'s may be replaced by the $\tilde
J_a^\mu(x)$'s:
\begin{eqnarray}
{\cal Z}_{BF}(\lambda)&=&\int{\cal D}B_\mu{\cal D}C_\mu
e^{-iS_{BF}[B,C]}\nonumber\\
&\times&e^{-i\int d^3x \left[
\tilde J_1^\mu(x)B_\mu(x)+
\tilde J_2^\mu(x)C_\mu(x)
\right]}
\end{eqnarray}
Due to the fact that the BF fields $B_\mu$ and $C_\mu$ are just dummy
field configurations over which a path integration is performed,
it is possible to rename them $\tilde B_\mu$ and $\tilde
C_\mu$ respectively. In conclusion, starting from the partition
function ${\cal 
  Z}_{BF}(\lambda)$ 
of Eq.~(\ref{zlambdawotilde}),
the partition function $\tilde {\cal
  Z}_{BF}(\lambda)$ appearing in Eq.~(\ref{maintop}) has been
recovered. In other words, it has 
been shown that:   
\begin{equation}
{\cal
  Z}_{BF}(\lambda)=
\tilde {\cal
  Z}_{BF}(\lambda)
\end{equation}
and thus in the identity (\ref{maintop}) we can replace $\tilde{\cal
  Z}_{BF}(\lambda)$ by ${\cal
  Z}_{BF}(\lambda)$:
\begin{equation}
\delta(m-\chi(\Gamma_1,\Gamma_2))=\int_{-\infty}^{+\infty}d\lambda
e^{-i\lambda m}{{\cal Z}}_{BF}(\lambda)\label{maintoptwo}
\end{equation}
This is the desired final result. 
Thanks to Eq.~(\ref{maintoptwo}), it will be possible to transform
the path
integral over all conformations of the $2s-$plat, which is complicated by
the cumbersome presence of the dirac delta function containing the
Gauss linking number, into a path integral over the
trajectories of a system
of $2s$ particles interacting with
magnetic fields.

In order to establish the analogy between polymers and anyons, which
will the subject of the next Section, it will be
convenient to quantize the BF model in the Coulomb
gauge\footnote{A similar approach like that proposed here can be found in
  \cite{froehlichking}.
In Ref.~\cite{froehlichking} the
 plats are however static, they do not fluctuate, and the light-cone
 gauge has been used.}:
\begin{equation}
\partial_{i}B^{i}=\partial_{i}C^{i}=0 \label{GG}
\end{equation}
After the gauge choice (\ref{GG}), the action of the BF model
(\ref{bfmodelaction}) becomes:
\begin{equation}
S_{BF,CG}[B,C]=\frac
{\kappa}{4\pi}\int\! d^{3}x[B_3
\varepsilon^{ij}\partial_{i}C_{j}+C_3
\varepsilon^{ij}\partial_{i}B_{j}] \label{CSCG}
\end{equation}
with $\varepsilon^{ij}=\varepsilon^{ij3}$ being the two-dimensional
completely antisymmetric tensor.
The gauge fixing term vanishes in the pure Coulomb gauge where the
conditions \ceq{GG} are strictly satisfied. Also the Faddeev-Popov term,
which in principle should be 
present in Eq.~\ceq{CSCG}, may be neglected because the ghosts
decouple from all other fields. Moreover, 
the requirement of transversality of (\ref{GG}) in the spatial directions
implies that the spatial components $B_{i}$ and $C_{i}$ 
of the BF fields
may be expressed in terms of
two scalar fields $b$ and $c$ via the Hodge decomposition:
\begin{equation}
B_{i}=\varepsilon_{ij}\partial^{j}b\qquad\qquad
C_{i}=\varepsilon_{ij}\partial^{j}c \label{hodgedef}
\end{equation}
After performing the above substitutions of fields in the BF action of
Eq.~\ceq{CSCG}, we obtain: 
\begin{equation}
S_{BF,CG}=\frac
{\kappa}{4\pi}\int d^{3}x[B_3\Delta c+C_3\Delta b] \label{bcaction}
\end{equation}
Let's compute now the propagator of the BF fields:
\begin{equation}
G_{\mu\nu}(\mathbf x,t;\mathbf
y,t^{\prime})=\langle B_{\mu}(\mathbf x,t),C_{\nu}(\mathbf
y,t^{\prime})\rangle
\end{equation}
From Eq.~\ceq{CSCG} it turns out that only the following components of the
propagator are different from zero:
\begin{equation}
G_{3i}(\mathbf x,t;\mathbf y,t^{\prime})
=\frac{\delta(t-t^{\prime})}{2\kappa}\epsilon_{ij}
\partial_{\mathbf y}^{j}\log|
\mathbf x-\mathbf y|^{2} \label{propone}
\end{equation}
\begin{equation}
G_{i3}(\mathbf x,t;\mathbf
y,t^{\prime})=-G_{3i}(
\mathbf x,t;\mathbf y,t^{\prime}) \label{proptwo}
\end{equation}
The path integration
over the scalar
fields 
$b$ and $c$
in
the partition function ${\cal Z}_{BF}(\lambda)$ 
is gaussian and could be in principle performed.
A natural question that arise at this point is the interpretation
of the topological constraint
\ceq{topconst} in the Coulomb gauge? As a matter of fact, the
BF propagator in the Coulomb gauge breaks explicitly the invariance of
the BF model under general three dimensional transformation.
It seems thus
hard to recover the form (\ref{ln}) of the Gauss
linking number in 
this gauge.
Of course, an
equivalent constraint should be obtained in the Coulomb gauge due to
gauge invariance. 
In Appendix B it will be shown by a direct calculation 
in the case
of a $4-$plat
that this is actually true. The computation of the expression of the
equivalent of the
Gauss linking number in the Coulomb gauge for a general $2s$-plat is
however technically complicated and will not be performed here.

\section{The partition function of a plat}\label{sectionIV}
In order to write the partition function of a $2s-$plat, we follow the
strategy explained in the previous
Section of dividing each trajectory $\Gamma_a$ into $2s_a$ open paths
$\Gamma_{a,I_a}$, $I_a=1,\ldots,2s_a$.
The statistical sum ${\cal Z}(\lambda)$ of the system,
that is performed over all
possible configurations $\mathbf r_{a,I_a}(t)$
of the subtrajectories $\Gamma_{a,I_a}$ using path integral methods,
is defined as follows:
\beq{
{\cal Z}(m)=\prod_{a=1}^2\prod_{I_a=1}^{2s_a}
\int_{\mbox{\scriptsize boundary}\atop{\mbox{\scriptsize
    conditions}}} \!{\cal D}\mathbf
r_{a,I_a}(t_{a,I_a})e^{-(S_{free}+S_{EV})}
\delta\left(
m-\chi(\Gamma_1,\Gamma_2)
\right)
}
{ppm0}
In the above equation the boundary conditions on the trajectories
enforce the 
constraints
\ceq{bcondtwoa} and \ceq{bcondtwo}. 
The free part of the action $S_{free}$ is given by:
\begin{equation}
S_{free}=\sum_{a=1}^{2}
\sum_{I_a=1}^{2s_{a}}\int_{\tau_{a,I_a-1}}^{\tau_{a,I_a}}\!dt_{a,I_a}(-1)^{I_a-1} 
g_{a,I_a} 
\left|\frac{d \mathbf r_{a,I_a}(t_{a,I_a})}{t_{a,I_a}}\right|^{2}\label{sfree}
\end{equation}
The parameters $g_{a,I_a}$, with that $g_{a,I_a}>0$, are
proportional to the inverse of the Kuhn lengths of the trajectories
$\Gamma_{a,I_a}$. 
They are also related 
to the total lengths of the trajectories $\Gamma_{a,I_a}$ as it is
discussed in 
Appendix A.
Let us note that 
$S_{free}$ is a positive definite
functional despite the presence of the factors $(-1)^{I_a-1}$. 
This can be easily proved by performing inside
 $S_{free}$ the transformations of Eqs.~\ceq{transfone} and \ceq{transftwo}
from the $t_{a,I_a}$'s to the $z_{a,I_a}$ variables:
\beq{
S_{free}=\sum_{a=1}^2\sum_{I_a=1}^{2s_a}\int_{\tau_{a,I_a-1}}^{\tau_{a,I_a}}
dz_{a,I_a}g_{a,I_a}\left|
\frac{d\mathbf r_{a,I_a}(z_{a,I_A})}{dz_{a,I_a}}
\right|^2
}{}
It is now evident that
$S_{free}$ is either positive or, if the $\mathbf
r_{a,I_a}(z_{a,I_a})$'s are constants, equal to zero. 

Since we wish to stress the analogy with directed paths moving in a
random media, we have also
to introduce a 
contribution with short-range interactions coming from the integration
over the random noises \cite{Kardar}. This is the origin of the contribution
$S_{EV}$ to the total action in
Eq.~\ceq{ppm0}. $S_{EV}$ is of the form:
\begin{eqnarray}
S_{EV}&=&\sum_{I=1}^{2s_1}
\sum_{J=1}^{2s_2}
\int_{\tau_{1,I-1}}^{\tau_{1,I}}\!dt_{1,I}
\int_{\tau_{2,J-1}}^{\tau_{2,J}}\!dt_{2,J}(-1)^{I+J-2}V(\mathbf
r_{1,I}(t_{1,I}))-\mathbf r_{2,J}(t_{2,J}))\delta(t_{1,I}-t_{2,J})\nonumber
\\
&+&
\frac{1}{2}\sum_{a=1}^{2}\sum_{I_a=1}^{2s_{a}}\sum_{J_a=1\atop
J_a\neq
I_a}^{2s_{a}}\int_{
\tau_{a,I_a-1}}^{\tau_{a,I_a}}\!dt_{a,I_a}\int_{\tau_{a,J_a-1}}^{\tau_{a,J_a}}
\!dt_{a,J_a}(-1)^{I_a+J_a-2}
V(\mathbf
r_{a,I_a}(t_{a,I_a}))-\mathbf
r_{a,J_a}(t_{a,J_a}))\nonumber\\
&\times&\delta(t_{a,I_a}-t_{a,J_a})\label{SEV} 
\end{eqnarray}
In the right hand side of the above equation, the first part describes
the interactions between the monomers belonging to different loops,
while the second part takes into account the interactions between the
monomers of the same loop.
If the random noises are gaussianly distributed,
the two-body potential $V(\mathbf r)$ 
is of the form:
\begin{equation}
V(\mathbf r)\sim V_{0}\delta(\mathbf r) \label{twobody}
\end{equation}
with $V_{0}$ being a positive constant. 
The factors $(-1)^{I+J-2}$
and $(-1)^{I_a+J_a-2}$ 
appearing in  Eq.~\ceq{SEV} are necessary to make the interactions
repulsive. This can be easily proved by
passing to the variables $z_{a,I_a}$
using the transformations of Eqs.~\ceq{transfone} and \ceq{transftwo}.

As explained in the previous Section, 
the delta function $\delta(m-\chi(\Gamma_1,\Gamma_2))$ 
appearing in the partition function ${\cal Z}(m)$ of Eq.~(\ref{ppm0}) may be
simplified by introducing
the BF-fields $B_\mu,C_\mu$ with action
\begin{equation}
 S_{BF}=
\frac{\kappa}{4\pi}\int\!{d^{3}x\varepsilon^{\mu\nu\rho}
B_{\mu}\partial_{\nu}C_{\rho}} \label{CSaction}
\end{equation} 
After performing the Fourier transform of the delta function according
to Eq.~(\ref{deltaft}) and exploiting Eq.~(\ref{maintoptwo}), the
partition function ${\cal Z}(m)$ becomes:
\beq{
{\cal Z}(m)=\int_{-\infty}^{+\infty} d\lambda e^{-i\lambda m}{\cal
  Z}(\lambda) 
}{ggg}
where
\beq{{\cal Z}(\lambda)=\int\!{\cal D}B_\mu{\cal D}C_\nu
e^{- i S_{BF}[B,C]}\prod_{a=1}^2\prod_{I_a=1}^{2s_a}
\int_{\mbox{\scriptsize boundary}\atop{\mbox{\scriptsize
    conditions}}} \!{\cal D}\mathbf r_{a,I_a}(t_{a,I_a})e^{-S}
}
{ppm}
The polymer action $S$ can be split as follows:
\begin{equation}
S=S_{free}+S_{EV}+S_{top}\label{actionsplitting}
\end{equation}
The expressions of $S_{free}$ and $S_{EV}$ are provided in Eqs.~(\ref{sfree})
and (\ref{SEV}) respectively.
Finally, using Eq.~\ceq{zlambdawotilde}, it is possible to realize
that the
topological contribution  $S_{top}$
to the polymer action turns out to be:
\begin{equation}
S_{top}=i\int d^3x\left[
J_1^\mu(x)B_\mu(x)+
J_2^\mu(x)C_\mu(x)
\right]
\label{stop}
\end{equation}
where the currents $J_a^\mu(x)$, $a=1,2$, are given in
Eq.~\ceq{currentwotilde}.

\section{An anyon field theory formulation of polymeric
  $2s-$plats}\label{sectionV} 
The starting point in this Section is the polymer statistical sum
${\cal Z}(\lambda)$ of Eq.~(\ref{ppm}). This is formally
equivalent to the partition function of a multicomponent
system of  anyon
particles.
To write this partition function in terms of fields, we  have to
perform an integration over all polymer trajectories $\mathbf
r_{a,I_a}(t_{a,I_a})$, 
$a=1,2$ and $I_a=1,\ldots,2s_{a}$. 
The passage to the field theoretical formulation is not just a formal
step, it allows to 
describe
the short range and topological interactions by means
a local and polynomial action. Before the introduction of
fields, these interactions are
both nonlocal and nonpolynomial.
The standard procedure 
in polymer physics
to pass from polymer trajectories to monomer densities and
thus to a field theory
consists in introducing
 auxiliary
fields.
In the case of the topological interactions, we have already seen that
the auxiliary fields are the BF fields
$B^\mu(x)$ and $C^{\mu}(x)$.
The short range interactions in Eq.~(\ref{SEV}) require instead
several scalar fields in order to be simplified.
The minimal number of these fields is
$2s_{a}+2s_{b}+2$. 
A couple of fields
$\phi_{1}(\mathbf x,t)$ and $\phi_{2}(\mathbf x,t)$ is needed
for the interaction between monomers belonging to
different loops.
The interactions between monomers belonging to the same loop
will be taken into account by the fields $\varphi_{1,I_a}(\mathbf
x,t)$'s and $\varphi_{2,J_b}(\mathbf x,t)$'s with
$I_a=1,\ldots,2s_{a}$, $J_b=1,\ldots,2s_{b}$.

The passages that lead to the final field theory are
well known in the polymer literature
\cite{FELAPLB1998,FELAJPA1999,FFNOVA,FFAnnPhys2002}.
After an integration over the auxiliary scalar fields
$\phi_{1}(\mathbf x,t),\phi_{2}(\mathbf x,t)$ and
$\varphi_{1,I_a}(\mathbf x,t),\varphi_{2,J_b}(\mathbf x,t)$, 
$I_a=1,\ldots,2s_{a}$, $J_b=1,\ldots,2s_{b}$, 
whose details
are explained in
Appendix C,
the expression of
the polymer partition function ${\cal Z}(\lambda)$ of Eq.~\ceq{ppm}
becomes:
\begin{eqnarray}
\!\!\!\!\!\!\!\!
{\cal Z}(\lambda)&=&\lim_{n_{1}\rightarrow 0 \atop n_{2}\rightarrow 0}
\int\!{\cal D}B_{\mu}{\cal D}C_{\mu}
\left[\prod_{I=1}^{2s_1}
\int\!{\cal D}\vec{\Psi}_{1,I} {\cal D}\vec{\Psi}_{1,I}^{*}
\psi_{1,I}^1(\mathbf r_{1,I}(\tau_{1,I-1}),\tau_{1,I-1})
\psi_{1,I}^{1*}(\mathbf r_{1,I}(\tau_{1,I}),\tau_{1,I})
\right]
 \nonumber\\
&&
\prod_{J=1}^{2s_{2}}
\int\!{\cal D}\vec{\Psi}_{2,J} {\cal D}\vec{\Psi}_{2,J}^*
\psi_{2,J}^1(\mathbf r_{2,J-1}(\tau_{2,J-1}),\tau_{2,J-1})
\psi_{2,J}^{1*}(\mathbf r_{2,J}(\tau_{2,J}),\tau_{2,J})
e^{-iS_{BF}}
e^{-{\cal A}}
e^{-{\cal A}_{EV}} \label{ppmthree}
\end{eqnarray}
where the $\vec{\Psi}_{a,I_{a}}^*$ and $\vec{\Psi}_{a,I_{a}}$, $a=1,2$,
$I_a=1,\ldots,2s_a$,
 are complex
replica fields:
\begin{equation}
\vec{\Psi}_{a,I_{a}}=(\psi_{a,I_{a}}^{1},\ldots,\psi_{a,I_{a}}^{n_{a}})\qquad
\vec{\Psi}_{a,I_{a}}^*=(\psi_{a,I_{a}}^{1*},\ldots,\psi_{a,I_{a}}^{n_{a*}})
\label{replicafielddef}
\end{equation}
The action ${\cal A}$ in Eq.~(\ref{ppmthree}) contains the free part
and the topological interactions:
\begin{eqnarray}
{\cal A}&=&\sum_{I=1}^{2s_{1}}\int_{\tau_{1,I-1}}^{\tau_{1,I}}\!dt(-1)^{I-1}
\int\!d^{2}\mathbf x\vec{\Psi}_{1,I}^{*}(\mathbf x,t)\cdot\nonumber \\
&&\left[\frac{\partial}{\partial t}-\frac{1}{4g_{1,I}}
\bigg(\boldsymbol \nabla -i\lambda(-1)^{I-1}\mathbf B(\mathbf x,t)\bigg)^{2}+
i\lambda(-1)^{I-1}B_3(\mathbf x,t)\right]
\vec \Psi_{1,I}(\mathbf x,t)\nonumber \\
&+&\sum_{J=1}^{2s_{2}}\int_{\tau_{2,J-1}}^{\tau_{2,J}}\!dt(-1)^{J-1}
\int\!d^{2}\mathbf x\vec{\Psi}_{2,J}^{*}(\mathbf x,t)\cdot\nonumber\\
&&\left[\frac{\partial}{\partial t}-\frac{1}{4g_{2,J}}
\bigg(\boldsymbol \nabla -\frac{i\kappa}{2\pi}(-1)^{J-1}\mathbf C(\mathbf
  x,t)\bigg)^{2}
+\frac{i\kappa}{2\pi}(-1)^{J-1}C_3(\mathbf x,t)\right]
\vec \Psi_{2,J}(\mathbf x,t) \label{afreetop}
\end{eqnarray}
The action ${\cal A}_{EV}$, given by
\begin{eqnarray}
{\cal A}_{EV}&=&\sum_{I,I'=1\atop I\ne I'}^{2s_1}\frac
{V_0}2\int_{\tau_{1,I-1}}^{\tau_{1,I}}
dt
\int_{\tau_{1,I'-1}}^{\tau_{1,I'}}
dt'\delta(t-t')\int d^2\mathbf x(-1)^{I+I'-2}\left|
\vec\Psi_{1,I}(\mathbf x,t)
\right|^2
\left|\vec\Psi_{1,I'}(\mathbf x,t')
\right|^2
\nonumber\\
&+&
\sum_{J,J'=1\atop J\ne J'}^{2s_2}\frac
{V_0}2\int_{\tau_{2,J-1}}^{\tau_{2,J}}
dt
\int_{\tau_{2,J'-1}}^{\tau_{2,J'}}
dt'\delta(t-t')\int d^2\mathbf x(-1)^{J+J'-2}\left|
\vec\Psi_{2,J}(\mathbf x,t)
\right|^2
\left|\vec\Psi_{2,J'}(\mathbf x,t')
\right|^2\nonumber\\
&+&
V_0\sum_{I=1}^{2s_{1}}\sum_{J=1}^{2s_{2}}
\int_{\tau_{1,I-1}}^{\tau_{1,I}}\!dt
\int_{\tau_{2,J-1}}^{\tau_{2,J}}dt^{\prime}\int \!d^{2}\mathbf x
(-1)^{I+J-2}
|\vec \Psi_{1,I}(\mathbf x,t)|^{2}|\vec
\Psi_{2,J}(\mathbf x,t^{\prime})|^{2}\delta(t-t^{\prime})
\label{aev}
\end{eqnarray}
is the analog of the action $S_{EV}$ written in the language of second
quantized fields and describes the short-range interactions.
Looking at Eqs.~\ceq{ppmthree}-\ceq{aev}, we see that we have succeeded
in our task, i.~e. 
the original polymer partition function \ceq{ppm} has been transformed
in an anyon field theory. 
The action ${\cal A}$ is formally equivalent to the action of a
multicomponent 
system of anyons subjected to the Coulomb interactions described by
${\cal A}_{EV}$. 
Similar systems have been discussed in connection with the
fractional quantum Hall effect and high $T_{C}$ superconductivity
\cite{wilczek2}.  
The only differences are the boundaries of the integrations over the
time, which in 
the present case depend on the heights of the points of maxima and
minima of the two trajectories $\Gamma_1,\Gamma_2$
and the fact that here the quasiparticles
are bosons of spin $n_{1}$ or $n_{2}$ considered in the limit
$n_{a}\rightarrow 0$,  $a=1,2$.

\section{Self-duality of the two-polymer problem}\label{sectionVI}
In this section we restrict ourselves for simplicity to 
$4-$plats. The partition function of a $4-$plat formed by two linked
polymers is obtained 
by putting $s_{1}=s_{2}=1$ in the
general
 partition function of a $2s-$plat given in
Eq.~\ceq{ppmthree}. Accordingly, the action ${\cal A}$ of Eq.~\ceq{afreetop}
 becomes in this particular case:
\begin{eqnarray}
{\cal A}&=&\int_{\tau_{1,0}}^{\tau_{1,1}}\!dt
\int\!d^{2}\mathbf x
\Bigg\lbrace\vec{\Psi}_{1,1}^{*}
\left[\frac{\partial}{\partial t}-\frac{1}{4g_{1,1}}\mathbf
  D^{2}(-\lambda,\mathbf B)+i\lambda
  B_3\right]\vec{\Psi}_{1,1}\nonumber\\ 
&+&\vec{\Psi}_{1,2}^{*}\left[\frac{\partial}{\partial
    t}-\frac{1}{4g_{1,2}}
\mathbf D^{2}(\lambda,\mathbf B)
-i\lambda B_{3}\right]\vec{\Psi}_{1,2}\Bigg\rbrace
\nonumber \\
&+&
\int_{\tau_{2,0}}^{\tau_{2,1}}\!dt\int d^2\mathbf x
\vec{\Psi}_{2,1}^{*}
\Bigg\lbrace\left[
\frac{\partial}{\partial t}
-\frac{1}{4g_{2,1}}
\mathbf D^{2}\left(
-\frac{\kappa}{2\pi}
,\mathbf C\right)
+\frac{i\kappa}{2\pi}C_3\right]\vec{\Psi}_{2,1}
\nonumber \\
&+&
\vec{\Psi}_{2,2}^{*}
\left[
\frac{\partial}{\partial t}
-\frac{1}{4g_{2,2}}
\mathbf D^{2}
\left(\frac{\kappa}{2\pi},\mathbf C\right)
-\frac{i\kappa}{2\pi}C_3
\right]
\vec{\Psi}_{2,2}\Bigg\rbrace \label{afourplat}
\end{eqnarray}
In writing the above equation, we have used the notations:
\begin{equation}
\mathbf D(\pm \lambda,\mathbf B)=\boldsymbol \nabla \pm i\lambda \mathbf B 
\qquad\qquad
\mathbf D\left(\pm \frac{\kappa}{2\pi},\mathbf C\right)=\boldsymbol \nabla \pm
i\frac{\kappa}{2\pi} \mathbf C 
\end{equation}
The short-range interaction term ${\cal A}_{EV}$ of Eq.~(\ref{aev}) 
simplifies in the case of a 4-plat  as follows:
\begin{eqnarray}
{\cal A}_{EV}&=&\sum_{I,J=1}^{2}V_{0}
\int_{\max[\tau_{1,0},\tau_{2,0}]}^{\min[\tau_{1,1},\tau_{2,1}]}\!dt 
\int\!d^{2}\mathbf x|\vec \Psi_{1,I}(\mathbf x,t)|^{2}|\vec
\Psi_{2,J}(\mathbf x,t)|^{2}\nonumber \\ 
&+&
\sum_{I\neq I'=1}^{2}\frac{V_{0}}{2}
\int_{\tau_{1,0}}^{\tau_{1,1}}\!dt
\int\!d^{2}\mathbf x|\vec \Psi_{1,I}(\mathbf x,t)|^{2}|\vec
\Psi_{1,I'}(\mathbf x,t)|^{2}\nonumber\\
&+&
\sum_{J\neq J'=1}^{2}\frac{V_{0}}{2}
\int_{\tau_{2,0}}^{\tau_{2,1}}\!dt
\int\!d^{2}\mathbf x|\vec \Psi_{2,J}(\mathbf x,t)|^{2}|\vec
\Psi_{2,J'}(\mathbf x,t)|^{2}\label{aev4plat}
 \end{eqnarray}
Finally, the BF contribution $iS_{BF}$ defined
in Eq.~\ceq{CSaction} remains unchanged.

The next goal is to find the classical field configurations which
minimize  the 
the 
energy ${\cal F}$ of the two-polymer system.
A sketchy derivation of these configurations can be found in
Ref.~\cite{FFPLA2004}. In the following, we will provide the details
that were missing in \cite{FFPLA2004}.
The energy $\cal F$
is given by:
\begin{equation}
 {\cal F}=iS_{BF}+{\cal A}+{\cal A}_{EV}
\end{equation}
where the expressions of $S_{BF}$, $\cal A$ and ${\cal A}_{EV}$ are
defined in Eqs.~(\ref{CSaction}), (\ref{afourplat}) and (\ref{aev4plat})
respectively.
To simplify the task of its minimization,
the short-range interactions will be neglected putting $V_0=0$ in
Eq.~\ceq{aev4plat}, so that 
${\cal A}_{EV}=0$.
This approximation is valid for instance for  polymer solutions which
are at the theta point. 
To proceed, we notice that the third components 
$B_3$ and $C_3$
of the BF
fields
play the role of pure Lagrange multipliers. Thus, they
can be integrated out from the 
partition function \ceq{ppmthree} giving as a result the following
constraints: 
\begin{eqnarray}
{\cal B}&=&2(|\vec \Psi_{2,1}|^{2}-|\vec \Psi_{2,2}|^{2})
\theta(\tau_{2,1}-t)\theta(t-\tau_{2,0}) \label{constrone}
\\
 {\cal C}&=&\frac{4\pi \lambda}{\kappa}(|\vec \Psi_{1,1}|^{2}-|\vec
 \Psi_{1,2}|^{2}) 
\theta(\tau_{1,1}-t)\theta(t-\tau_{1,0})  \label{constrtwo}
\end{eqnarray}
where ${\cal B}$ and $\cal C$ are the magnetic fields associated to
the vector potentials $B_{i}$ and $C_{i}$ respectively: 
\begin{equation}
 {\cal
   B}=\partial_{1}B_{2}-\partial_{2}B_{1}=
\varepsilon^{ij}\partial_{i}B_{j} \label{magb}  
\end{equation}
\begin{equation}
{\cal
  C}=\partial_{1}C_{2}-\partial_{2}C_{1}=
\varepsilon^{ij}\partial_{i}C_{j} \label{magc} 
\end{equation}
In Eqs.~(\ref{constrone})  and (\ref{constrtwo}),
$\theta (t)$ denotes the Heaviside function $\theta (t)=0$ if $t<0$ and
$\theta (t)=1$ if $t\ge 0$. 
We will look here only for static field configurations, i.e. those
which satisfy the relations: 
\begin{equation}
\frac{\partial}{\partial t}\psi_{a,I_{a}}^{\sigma_{a}}=
\frac{\partial}{\partial t}\psi_{a,I_{a}}^{*\sigma_{a}}=0  \label{statereq}
\end{equation}
for all values of $a=1,2$ $I_{a}=1,2$ and $\sigma_{a}=1,\ldots,n_{a}$,
where the $n_{a}$'s denote the numbers of replicas. 
To avoid problems with the presence of the Heaviside functions in the
expression of the magnetic 
fields, we will assume  that 
\beq{\tau_{1,0}=\tau_{2,0}\equiv\tau_{0} \qquad\mbox{and}\qquad
\tau_{1,1}=\tau_{2,1}=\tau_{1}}{assumption}
In this way, the parameter $t$, whose range is changing depending on which
subtrajectory is parametrized, is always be defined in the interval
$[\tau_{0},\tau_{1}]$ as a real time. 
At this point,  the static
energy ${\cal F}_{st}$ may be written as follows: 
\begin{eqnarray}
{\cal F}_{st}&=&(\tau_1-\tau_0)\int d^{2}\mathbf x
\left[\frac{1}{4g_{1,1}}\left|\mathbf D(-\lambda,\mathbf B)\vec
  \Psi_{1,1}\right|^{2}+ 
\frac{1}{4g_{1,2}}
\left|\mathbf D(\lambda,\mathbf B)\vec \Psi_{1,2}\right|^{2}\right]\nonumber \\
&+&(\tau_{1}-\tau_{0})\int d^{2}\mathbf x
\left[ \frac{1}{4g_{2,1}}\left |\mathbf D\left(-\frac{\kappa}{2\pi},\mathbf C
\right)\vec \Psi_{2,1}\right|^{2}+
\frac{1}{4g_{2,2}}\left|\mathbf D\left(\frac{\kappa}{2\pi},\mathbf
C\right)\vec 
\Psi_{2,2}\right|^{2}\right] 
\label{staticfe}
\end{eqnarray}
The vector potentials $\mathbf B$ and $\mathbf C$ in the above
equations are determined by the relations (\ref{constrone}--\ref{magc}).
The analogy with the anyon problem suggests the application of the
Bogomol'nyi identities \cite{dunne}.  
For a single theory of complex scalar fields $\psi^{*},\psi$ minimally
coupled to an abelian gauge field $\mathbf a$, 
these identities look as follows:
\begin{equation}
|\mathbf D(\gamma,\mathbf a)\psi|^{2}=
|D_{\pm}(\gamma,\mathbf a)\psi|^{2}\mp \gamma {\tt b}|\psi|^{2}\pm
\varepsilon^{ik}\partial_{i}j_{k} \label{bogomi} 
\end{equation}
where $\mathbf D(\gamma, \mathbf a)=\boldsymbol\nabla-i\gamma
\mathbf a$ is the covariant derivative and
\begin{equation}
D_{\pm}(\gamma, \mathbf a)=
D_{1}(\gamma, \mathbf a)\pm iD_{2}(\gamma, \mathbf a)
\end{equation}
Here $D_{i}(\gamma,a)$, $i=1,2$, denotes the components of $\mathbf
D(\gamma, \mathbf a)$, while
\begin{equation}
 {\tt b}=\partial_1a_{2}-\partial_{2}a_{1}
\end{equation}
is the magnetic field.
Finally
\begin{equation}
j_k=\frac{1}{2i}[\psi^{*}D_k(\gamma, \mathbf a)\psi -\psi
  D_k(\gamma, \mathbf a
)\psi^{*}] 
\end{equation}
is the current related to the abelian gauge group of symmetry.
Let us notice that the term in Eq.~\ceq{bogomi} 
containing $j_k$
is a total derivative,
so that it can be omitted in our case, in which the space has no boundaries.

Coming back to the problem of minimizing the static free energy of
Eq.~\ceq{staticfe}, we can 
now apply the Bogomol'nyi 
identities for all $2n_1+2n_2$ replica fields.  Actually,
Eq.~\ceq{bogomi} defines two different identities, 
depending on the choice of sign. This fact may be used to simplify the
calculations. In particular, we will choose 
the $+$ sign when the scalar fields are coupled to $\mathbf B$ and the
$-$ sign when the scalar fields are coupled to $\mathbf C$.
As a result we obtain:
\beq{{\cal F}_{st}={\cal F}_1+{\cal F}_2}{newstaticfe}
with
\begin{eqnarray}
{\cal F}_1&=&(\tau_1-\tau_0)\int d^{2}\!\mathbf x\Bigg\lbrace
\frac{1}{4g_{1,1}}\left|\mathbf D_{+}(-\lambda,\mathbf B)\vec \Psi_{1,1}\right|^{2}+
\frac{1}{4g_{1,2}}
\left|\mathbf D_{+}(\lambda,\mathbf B)\vec \Psi_{1,2}\right|^{2}\nonumber \\
&+&
\frac{1}{4g_{2,1}}\left|\mathbf D_-\left(-\frac{\kappa}{2\pi},\mathbf
C\right)\vec \Psi_{2,1}\right|^{2}+ 
\frac{1}{4g_{2,2}}\left|\mathbf D_-\left(\frac{\kappa}{2\pi},\mathbf
C\right)\vec \Psi_{2,2}
\right|^{2}
\Bigg\rbrace\label{newstaticfeone}
\end{eqnarray}
\begin{eqnarray}
{\cal F}_2&=&(\tau_1-\tau_0)\int d^{2}\!\mathbf x\Bigg\lbrace
\frac{\lambda}{4g_{1,1}}{\cal B}\left|\vec \Psi_{1,1}\right|^{2}-
\frac{\lambda}{4g_{1,2}}{\cal B}\left|\vec \Psi_{1,2}\right|^{2}
-\frac{\kappa}{8\pi g_{2,1}}{\cal C}\left|\vec \Psi_{2,1}\right|^{2}
\nonumber\\
&+&
\frac{\kappa}{8\pi g_{2,2}}{\cal C}\left|\vec \Psi_{2,2}\right|^{2}
\Bigg\rbrace
\label{newstaticfetwo}
\end{eqnarray}
If we substitute in Eq.~\ceq{newstaticfetwo}
the expressions of the magnetic fields
 ${\cal B}$ and ${\cal C}$ in terms of the replica fields given by
Eqs.~\ceq{constrone}-\ceq{constrtwo},
it is easy to realize that ${\cal F}_2$ becomes of the form: 
\begin{eqnarray}
{\cal F}_2&=&\frac\lambda 2(\tau_1-\tau_0)\!\int d^{2}\mathbf x
\left[ \frac 1{g_{1,1}}\left(|\vec \Psi_{2,1}|^{2}-|\vec \Psi_{2,2}|^{2}\right)
|\vec\Psi_{1,1}|^2
-
\frac 1{g_{1,2}}\left(|\vec \Psi_{2,1}|^2-|\vec \Psi_{2,2}|^2\right)
|\vec \Psi_{1,2}|^2\right.
\nonumber\\
&-&\left.\frac 1{g_{2,1}}\left(|\vec \Psi_{1,1}|^2-|\vec \Psi_{1,2}|^2\right)
|\vec \Psi_{2,1}|^2
+\frac 1{g_{2,2}}\left(|\vec \Psi_{1,1}|^2-
|\vec \Psi_{1,2}|^2\right)|\vec \Psi_{2,2}|^2\right]
\label{selfdual}
\end{eqnarray}
It turns out from the above equation
that the presence of the topological constraints induces
changes 
in the energy of two linked polymers which
consists in the appearance of short-range
interactions with coupling constants
proportional to
\begin{equation}
\frac{\lambda}{g_{a,I_a}}(\tau_1-\tau_0)\label{factor}
\end{equation}
These interactions clearly interfere with the short-range
interactions given in Eq.~(\ref{aev4plat}), 
which have  potentials of the same structure, characterized by
fourth-order powers of the fields, 
but have different coupling constants. In particular, in
Eq.~(\ref{aev4plat}) the coupling constant $V_0$ is always positive,
while the coupling constants in Eq.~(\ref{factor}) can be either
positive or negative.
This shows that the topological constraints have
nontrivial effects on the short-term interactions acting on the
monomers. These effects
have been already observed in
experiments, see for example Ref.~\cite{levene}. Analytically, 
the influence of the topological constraints has been quantitatively
described using various approximations \cite{ferrari,vilgisotto,
  otto}. Thanks to the analogy between anyons and $2s-$plats
established here, we have
been able to derive Eq.~(\ref{selfdual}),
which represents a direct confirmation at a nonperturbative level
of the appearance  of
interactions associated to  topological constraints.


Let us now go back to the expression of the static energy ${\cal F}_{st}$ of
Eq.~\ceq{newstaticfe}. Looking at the form of its components ${\cal F}_1$ and
${\cal F}_2$ of Eqs.~(\ref{newstaticfeone}) and (\ref{selfdual}),
it is possible to conclude that
${\cal F}_{st}$ 
 is formally equivalent to the Hamiltonian of a set of complex
scalar fields coupled to the BF fields $B_{\mu}$ and
$C_{\mu}$. This kind of theory is known to have self-dual solutions
\cite{wilczek2,dunne}.  
%
In general, the search of self-dual solutions is not a simple task,
because of the non-linear character of the
classical equations of motion. Up to now, this problem has been 
 solved in general only using numerical methods.
Despite these difficulties, however, 
it is still possible to investigate
the self-dual point
analytically by restricting ourselves
to the region of the space of physical parameters 
 in which the attractive and repulsive forces
 appearing in ${\cal F}_2$ counterbalance themselves. 
In the present context, the self-duality is achieved when the following
conditions are satisfied: 
\begin{equation}
g_{1,1}=g_{1,2}=g_{2,1}=g_{2,2}=g \label{selfdualcond}
\end{equation}
If the equalities in Eq.~\ceq{selfdualcond} are valid, in fact, the
potentials in the right hand side of Eq.~\ceq{selfdual} 
vanish identically. As a consequence, the energy
\ceq{newstaticfe} becomes self-dual, i.e. it can be written as a sum
of 
self-dual contribution:
\begin{eqnarray}
 {\cal F}_{st}&=&\frac{(\tau_1-\tau_0)}{4g}\int d^{2}\mathbf x
\left[\left|\mathbf D_{+}(-\lambda,\mathbf B)\vec \Psi_{1,1}\right|^2+
\left|\mathbf D_+(\lambda,\mathbf B)\vec \Psi_{1,2}\right|^2\right]\nonumber \\
&+&\frac{(\tau_1-\tau_0)}{4g}\int\!d^{2}\mathbf x
\left[
\left|\mathbf D_-\left(-\frac{\kappa}{2\pi},\mathbf C \right)\vec
\Psi_{2,1}\right|^2 
+
\left|\mathbf D_-\left(\frac{\kappa}{2\pi},\mathbf C\right)\vec \Psi_{2,2}\right|^2\right]
\label{sdstfe}
\end{eqnarray}
In anyon field theories the self-duality condition \ceq{selfdualcond}
has a very physical meaning, see for example \cite{dunne}, but
its interpretation
in the case of the $2s-$plat is much more difficult.
Certainly  the self-duality condition (\ref{selfdualcond})
is related both to the length and rigidity of the
polymer trajectories.
Indeed,  the
parameters $g_{a,I_a}$  can be identified with the inverse
 of the Kuhn lengths
of the subtrajectories $\Gamma_{a,I_a}$ and thus determine their
rigidity. Moreover, in Appendix A it is shown how the lengths of the
subtrajectories depend on the $g_{a,I_a}$'s, see
Eq.~\ceq{pollength}. Therefore, it is clear that the relations
\ceq{selfdualcond}
are also imposing conditions on the lengths of the trajectories 
$\Gamma_{1,1}$,$\Gamma_{1,2}$,$\Gamma_{2,1}$ and $\Gamma_{2,2}$,
which must have in the average the same lengths in order to attain the
self-dual point.

In the next Section we will derive some explicit self-dual configurations
which minimize the free energy ${\cal F}_{st}$ of Eq.~\ceq{sdstfe}.
\section{Self-dual solutions of the two-polymer problem}\label{sectionVII}
The task of this Section is to find classical solutions of the
equations of motion which minimize the energy ${\cal F}_{st}$ of
Eq.~\ceq{sdstfe}. 
The classical equations of motion
read as follows:
\begin{eqnarray}
\mathbf D_{+}(-\lambda,\mathbf B)\psi_{1,1}^{\sigma_{1}}&=&0\label{cleq1}\\
\quad \mathbf D_{+}(\lambda,\mathbf B)\psi_{1,2}^{\sigma_{1}}&=&0\label{cleq2}\\
\quad\mathbf D_{-}\left(-\frac{\kappa}{2\pi},\mathbf C
\right)\psi_{2,1}^{\sigma_2}&=&0 
\label{cleq3}\\
\mathbf D_{-}\left(-\frac{\kappa}{2\pi},\mathbf
C\right)\psi_{2,2}^{\sigma_{2}}&=&0\label{cleq4} 
\end{eqnarray}
$\sigma_{1}$,$\sigma_{2}$ being replica indexes.
To Eqs.~(\ref{cleq1}--\ref{cleq4}) one should add the constraints
\ceq{constrone} and \ceq{constrtwo}:
\begin{eqnarray}
\epsilon^{ij}\partial_iB_j
&=&2\left(|\vec \Psi_{2,1}|^2-
|\vec \Psi_{2,2}|^2\right)
\theta(\tau_1-t)\theta(t-\tau_0)\label{cstr1} \\
\epsilon^{ij}\partial_{i}C_{j}
&=&\frac{4\pi \lambda}{\kappa}(|\vec \Psi_{1,1}|^2-
|\vec \Psi_{1,2}|^2)
\theta(\tau_1-t)\theta(t-\tau_0)\label{cstr2}
\end{eqnarray}
To avoid analytical complications due to the presence of the Heaviside
theta functions, we have assumed as in Eq.~\ceq{assumption} that
$\tau_{1,0}=\tau_{2,0}=\tau_0$ and $\tau_{1,1}=\tau_{2,1}=\tau_1$.
Moreover, in the following
we will restrict ourselves to the replica symmetric solutions
of Eqs.~(\ref{cleq1}--\ref{cleq4}) and
(\ref{cstr1}--\ref{cstr2}) by putting:
\begin{eqnarray}
\psi_{1,I}^{\sigma_{1}}&=&\psi_{1,I}\quad \mbox{for} \quad1\leq \sigma_{1}\leq n_{1}\quad  I=1,2\nonumber\\
\psi_{2,J}^{\sigma_{2}}&=&\psi_{2,J} \quad \mbox{for} \quad1\leq \sigma_{2}\leq n_{2} \quad J=1,2
\end{eqnarray}
In this way, the explicit form of the equations of motion
(\ref{cleq1}--\ref{cleq4}) and of the constraints
(\ref{cstr1}--\ref{cstr2}) looks as follows:
\begin{eqnarray}
\left[\partial_1-i\lambda B_1+i\left(\partial_2-i\lambda B_2\right)\right]\psi_{1,1}&=&0
\label{eqexp1}\\
\left[\partial_1+i\lambda B_1+i\left(\partial_2+i\lambda B_2\right)\right]\psi_{1,2}&=&0
\label{eqexp2}\\
\left[\partial_1-\frac{i\kappa}{2\pi} C_1-i\left(\partial_2-\frac{i\kappa}{2\pi} C_2\right)\right]\psi_{2,1}&=&0
\label{eqexp3}\\
\left[\partial_1+\frac{i\kappa}{2\pi} C_1-i\left(\partial_2+\frac{i\kappa}{2\pi} C_2\right)\right]\psi_{2,2}&=&0
\label{eqexp4}
\end{eqnarray}
\begin{eqnarray}
\epsilon^{ij}\partial_iB_j&=&2n_1\left(| \psi_{2,1}|^2-
| \psi_{2,2}|^2\right)\label{cstrexp1} \\
\epsilon^{ij}\partial_{i}C_{j}&=&\frac{4n_2\pi \lambda}{\kappa}
\left(|\psi_{1,1}|^2-|\psi_{1,2}|^2\right)\label{cstrexp2}
\end{eqnarray}
At this point we pass to polar coordinates by performing the
transformations:
\begin{equation}
\psi_{a,I_a}=e^{i\omega_{a,I_a}}\rho_{a,I_a}^{1/2}\label{trsf1}
\end{equation}
After the above change of variables in 
Eqs.~(\ref{eqexp1}--\ref{cstrexp2}),
we obtain by separating the real and imaginary parts:
\begin{eqnarray}
\partial_1\omega_{1,1}-\lambda B_1+\frac{1}{2}\partial_2\log\rho_{1,1}&=&0
\label{eqpc1}\\
-\partial_2\omega_{1,1}+\lambda B_2+\frac{1}{2}\partial_1\log\rho_{1,1}&=&0
\label{eqpc2}\\
\partial_1\omega_{1,2}+\lambda B_1+\frac{1}{2}\partial_2\log\rho_{1,2}&=&0
\label{eqpc3}\\
-\partial_2\omega_{1,2}-\lambda B_2+\frac{1}{2}\partial_1\log\rho_{1,2}&=&0
\label{eqpc4}\\
\partial_1\omega_{2,1}-\frac{\kappa}{2\pi} C_1-\frac{1}{2}\partial_2\log\rho_{2,1}&=&0
\label{eqpc5}\\
\partial_2\omega_{2,1}-\frac{\kappa}{2\pi} C_2+\frac{1}{2}\partial_1\log\rho_{2,1}&=&0
\label{eqpc6}\\
\partial_1\omega_{2,2}+\frac{\kappa}{2\pi} C_1-\frac{1}{2}\partial_2\log\rho_{2,2}&=&0
\label{eqpc7}\\
\partial_2\omega_{2,2}+\frac{\kappa}{2\pi} C_2+\frac{1}{2}\partial_1\log\rho_{2,2}&=&0
\label{eqpc8}
\end{eqnarray}
\begin{eqnarray}
\epsilon^{ij}\partial_iB_j&=&2n_1\left(\rho_{2,1}-
\rho_{2,2}\right)\label{cstrpc1} \\
\epsilon^{ij}\partial_{i}C_{j}&=&\frac{4n_2\pi \lambda}{\kappa}
\left(\rho_{1,1}-\rho_{1,2}\right)\label{cstrpc2}
\end{eqnarray}
To solve equations (\ref{eqpc1}--\ref{eqpc8}) with respects to the
unknowns $\omega_{a,I_a}$ and $\rho_{a,I_a}$, we proceed as follows.
First of all, we isolate from Eq.~(\ref{eqpc1}) and Eq.~(\ref{eqpc3})
the same quantity $\lambda B_1$. By requiring
that the expressions of $\lambda B_1$ provided by Eqs.~(\ref{eqpc1}) and
(\ref{eqpc3}) are equal, we obtain the consistency condition:
\begin{equation}
\partial_1\omega_{1,1}+\frac{1}{2}\partial_2\log\rho_{1,1}=
-\partial_1\omega_{1,2}-\frac{1}{2}\partial_2\log\rho_{1,2}\label{cstc1} 
\end{equation}
A possible solution of Eq.~(\ref{cstc1}) is:
\begin{equation}
\omega_{1,1}=-\omega_{1,2}\quad \mbox{and} \quad \rho_{1,1}=\frac{A_1}{\rho_{1,2}}\label{ansatz1}
\end{equation}
where $A_1$ is a constant factor.
As well, we could require that the two different expressions of the
quantity $\lambda B_2$ obtained from
Eqs.~(\ref{eqpc2}) and (\ref{eqpc4}) are equal. 
However, in this way one obtains once again the condition
(\ref{cstc1}),
%
which can be solved by applying the
ansatz (\ref{ansatz1}).
In a similar way, it is possible to extract from
equations 
(\ref{eqpc5}--\ref{eqpc8}) the conditions:
\begin{equation}
\omega_{2,1}=-\omega_{2,2}\quad \mbox{and} \quad \rho_{2,1}=\frac{A_2}{\rho_{2,2}}\label{ansatz2}
\end{equation}
with $A_2$ being a constant.\\
Thanks to (\ref{ansatz1}) and (\ref{ansatz2}),
the number of unknowns to be computed is reduced. For instance, if we
know the expressions of
$\omega_{1,1},\omega_{2,1},\rho_{1,1}$ and $\rho_{2,1}$,  the
classical field configurations $\omega_{1,2},\omega_{2,2},\rho(1,2)$ 
and $\rho_{2,2}$ can be derived using Eqs.~(\ref{ansatz1}) and
(\ref{ansatz2}).  
As a consequence, the
system of equations (\ref{eqpc1}--\ref{cstrpc2}) reduces to:
\begin{eqnarray}
\lambda
B_1&=&\partial_1\omega_{1,1}+\frac{1}{2}\partial_2\log\rho_{1,1}\label{eqred1}\\
\lambda
B_2&=&\partial_2\omega_{1,1}-\frac{1}{2}\partial_1\log\rho_{1,1}\label{eqred2}\\
\frac{\kappa}{2\pi}
C_1&=&\partial_1\omega_{2,1}-\frac{1}{2}\partial_2\log\rho_{2,1}\label{eqred3}\\
\frac{\kappa}{2\pi}
C_2&=&\partial_2\omega_{2,1}+\frac{1}{2}\partial_1\log\rho_{2,1}\label{eqred4}\\
\partial_1B_2-\partial_2
B_1&=&2n_1\left(\rho_{2,1}-\frac{A_2}{\rho_{2,1}}\right)\label{cstrred1}\\
\partial_1C_2-\partial_2 C_1&=&\frac{4n_2\pi\lambda}{\kappa}\left(\rho_{2,1}-\frac{A_2}{\rho_{2,1}}\right)\label{cstrred2}
\end{eqnarray}
where
we have used the fact that
$\epsilon^{ij}\partial_iB_j=\partial_1B_2-\partial_2B_1$ and
$\epsilon^{ij}\partial_iC_j=\partial_1C_2-\partial_2C_1$.
Eqs.~(\ref{cstrred2}) contain
only the unknowns $\omega_{1,1},\omega_{2,1},\rho_{1,1}$ and
$\rho_{2,1}$ that have still to be determined.

By subtracting term by term the two
equations resulting from the derivation of
Eqs.~(\ref{eqred1}) and (\ref{eqred2})
with respect to the variables $x^2$ 
and $x^1$ respectively, we obtain as an upshot the relation:
\begin{equation}
\lambda\left(\partial_1B_2-
\partial_2B_1\right)=\partial_1\partial_2\omega_{1,1}-
\partial_2\partial_1\omega_{1,1}-\frac{1}{2}\Delta\log\rho_{1,1}\label{sub12}
\end{equation}
with $\Delta=\partial_1^2+\partial_2^2$ being the two-dimensional
Laplacian.\\ 
Assuming that $\omega_{1,1}$ is a regular function satisfying the
relation 
\begin{equation}
\partial_1\partial_2\omega_{1,1}-\partial_2\partial_1\omega_{1,1}=0
\end{equation}
Eq.~(\ref{sub12}) becomes:
\begin{equation}
\lambda\left(\partial_1B_2-\partial_2B_1\right)=-\frac{1}{2}\Delta\log\rho_{1,1}\label{sub12nv}
\end{equation}
An analogous identity can be derived starting from Eqs.~(\ref{eqred3})
and (\ref{eqred4}): 
\begin{equation}
\frac{\kappa}{\pi}\left(\partial_1C_2-\partial_2C_1\right)=\Delta\log\rho_{2,1}\label{sub34nv}
\end{equation}
The compatibility of (\ref{sub12nv}) and (\ref{sub34nv}) with the
constraints (\ref{cstrred1}) and (\ref{cstrred2}) respectively leads
to the following conditions between $\rho_{1,1}$ and $\rho_{2,1}$:
\begin{eqnarray}
\Delta\log\rho_{1,1}&=&4\lambda
n_1\left(\frac{A_2}{\rho_{2,1}}-\rho_{2,1}\right)\label{cnstcy1}\\
\Delta\log\rho_{2,1}&=&4\lambda n_2\left(\rho_{1,1}-\frac{A_1}{\rho_{1,1}}\right)\label{cnstcy2}
\end{eqnarray}
The fact that $\rho_{1,1}$ and $\rho_{2,1}$
appear in a symmetric way in Eqs.~(\ref{cnstcy1}) and (\ref{cnstcy2}),
suggests the following ansatz:
\begin{equation}
\rho_{2,1}=\frac{A_3}{\rho_{1,1}}\label{ansatz3}
\end{equation}
$A_3$ being a constant.
It is easy to check that with this ansatz Eqs.~(\ref{cnstcy1}) and
(\ref{cnstcy2}) remain compatible provided:
\begin{equation}
\frac{A_2}{A_3} = -\frac{n_2}{n_1}\quad\mbox{and}\quad\frac{A_3}{A_1} = -\frac{n_2}{n_1}
\label{constconst}
\end{equation}
We choose $A_1$ to be the independent constant, while $A_2$ and $A_3$
are constrained 
by Eq.~\ceq{constconst} to be dependent on $A_1$:
\begin{equation}
A_2 = \left(\frac {n_2}{n_1}\right )^2 A_1\qquad A_3 = -\frac {n_2}{n_1}A_1
\end{equation}
We are now left only with the task of computing the explicit
expression of $\rho_{1,1}$.
This may be obtained by solving  
the equation:
\begin{equation}
\Delta\log{\rho_{1,1}} = 4\lambda n_2 \left(  \frac{A_1}{\rho_{1,1}}
-\rho_{1,1}   \right) 
\label{rho11fmal}
\end{equation}
The other quantities $\rho_{2,1}$, $\rho_{1,2}$ and $\rho_{2,2}$ can
be derived using  
the relations \ceq{ansatz3}, \ceq{ansatz1} and \ceq{ansatz2}
respectively.
Eq.~(\ref{rho11fmal}) may be cast in a more familiar form by putting:
$\eta=\ln\left(\frac{\rho_{1,1}}{\sqrt{A_1}}\right)$. After this
substitution, Eq.~(\ref{rho11fmal}) becomes the Euclidean sinh--Gordon
equation with respect to $\eta$:
\beq{\Delta\eta=8\lambda n_2\sqrt{A_1}\sinh\eta}{sinhgordon}
Next, it is possible to determine the magnetic fields $\mathbf B$ and
$\mathbf C$ from 
Eqs.~\ceq{cstrred1} and \ceq{cstrred2}. In the Coulomb gauge, in fact,
the two dimensional 
vector potentials $\mathbf B$ and $\mathbf C$ can be represented using
two scalar fields 
$b$ and $c$ as follows (see also Eq.~(\ref{hodgedef})):
\begin{equation}
\mathbf B = (-\partial_2 b, \partial_1 b) \qquad \mathbf C =
(-\partial_2 c, \partial_1 c) 
\label{hodgedec}
\end{equation}
Performing the above substitutions
in Eqs.~\ceq{cstrred1} and \ceq{cstrred2}, it turns out that $b$ and
$c$ satisfy the relations: 
\begin{equation}
\Delta b = 2n_2 (\rho_{1,1} - \frac {A_1}{\rho_{1,1}})
\label{eqb}
\end{equation}
\begin{equation}
\Delta c = \frac{4n_2\pi\lambda}{\kappa} (\rho_{1,1} - \frac
{A_1}{\rho_{1,1}}) 
\label{eqc}
\end{equation}
The solution of Eqs.~\ceq{eqb} and \ceq{eqc} can be easily derived
with the help  
of the method of the Green functions once the expression of
$\rho_{1,1}$ is known. 
Finally, the phases $\omega_{1,1}$, $\omega_{1,2}$, $\omega_{2,1}$ and
$\omega_{2,2}$ 
are computed using Eqs.~\ceq{eqred1}--\ceq{eqred4}.
In fact, remembering that we assumed that $\omega_{1,1}=-\omega_{1,2}$
and $\omega_{2,1} = \omega_{2,2}$ 
in \ceq{ansatz1} and \ceq{ansatz2} respectively, we have only to
determine $\omega_{1,1}$ and $\omega_{2,1}$. 
By deriving Eq.~\ceq{eqred1} with respect to $x^1$ and
Eq.~\ceq{eqred2} with respect to $x^2$, we obtain: 
\begin{eqnarray}
\lambda \partial_1B_1 &=& \partial_1^2 \omega_{1,1}  + \frac 12
\partial_1\partial_2\log{\rho_{1,1}}\nonumber\\ 
\lambda \partial_2B_2 &=& \partial_2^2 \omega_{1,1}  - \frac 12
\partial_2\partial_1\log{\rho_{1,1}} 
\end{eqnarray} 
On the other side, by adding term by term the above two equations and
using the fact that 
in the Coulomb gauge
the magnetic field $\mathbf B$ is completely transverse, it is
possible to show that:
\begin{equation}
\Delta \omega_{1,1} = 0 \label{fdsfsd1}
\end{equation}
Proceeding in a similar way with Eq.~\ceq{eqred3} and \ceq{eqred4} it
is possible to derive also the relation 
satisfied by $\omega_{2,1}$:
\begin{equation}
\Delta\omega_{2,1} = 0\label{fdsfsd2}
\end{equation}
\section{Conclusions}\label{sectionVIII}
In this work a system of two polymers forming a nontrivial link has
been considered. The topological properties of the link have been
described by using the Gauss linking invariant. This is a weak
topological invariant, but when applied to a $2s-$plat configuration,
which cannot be destroyed because the $2s$ points of maxima and minima
are kept fixed, its capabilities to distinguish the
changes of topology are greatly enhanced. The reason is the synergy
between the constraint imposed by the Gauss linking number
and those imposed by the fact that the
polymer system cannot 
escape the set of conformations allowed in a $2s-$plat.
We have also seen in Appendix B
what is the meaning of these constraints from the point of view of the
$2s-$plat. Basically, the sum of the winding numbers of all pairs of the
subtrajectories $\Gamma_{a,I_a}$ are constrained to be equal to some
integer multiple of $2\pi$. Moreover, since the endpoints of the
trajectories are fixed, also the winding number between two different
trajectories is fixed up to multiples of $2\pi$. Allowed are only the
topology changes such that an amount of the winding angle of two
subtrajectories  is transferred in units of $2\pi$ to the winding
angle of another couple of subtrajectories.
This result paves the way to a treatment of polymer knots or links
constructed from tangles. Polymers of this kind are relevant in
biochemistry because nontrivial knot configurations appearing as a major 
pattern in DNA rings are mostly in the form of tangles \cite{sumners}.

A crucial point of the connection shown in this paper between
$2s-$plats and anyons is the possibility
of elimininating
the cumbersome topological constraint (\ref{topconst}) from the
partition function ${\cal Z}(\lambda)$ of a $2s-$plat by
introducing BF fields.
Indeed,
the
delta function fixing the constraint  (\ref{topconst}) can be
represented using the Fourier transform of the amplitude of gauge
invariant and metric independent observables of an abelian BF model.
This has been proved in Eq.~(\ref{maintoptwo}). The proof of this
relation is not trivial because
the $2s$ subtrajectories in which
the original $2s-$plat has been split are open and parametrized by
a special variable, the parameter $t$, which is proportional to one
of the spatial components of the subtrajectories themselves.

Thanks to the identity (\ref{maintoptwo}) it has been possible to
interpret the problem of the statistical mechanics of a $2s-$plat
as that of a two-component anyon gas with $2s_1$ particles of kind 1 and
$2s_2$ particles of kind 2 interacting via short range potentials, see
Eqs.(\ref{ggg}--\ref{stop}).
The trajectories of the quasi-particles correspond in the polymer
analog to the $2s$ directed trajectories $\Gamma_{a,I_a}$, $a=1,2$,
$I_a=1,\ldots, 2s_a$, which are traversed by the fictitious currents
(\ref{currentwotilde}). 
The Gauss
linking number can be interpreted as the circulation of the magnetic
field generated by the current traversing the loop $\Gamma_1$ with
respect to the closed contour formed by the loop $\Gamma_2$
\cite{FFAnnPhys2002,Brereton}. 

The system of quasiparticles with partition function
Eq.~(\ref{ppm})  has been further
mapped into a two-component anyon field theory, whose final form is
displayed in Eqs.~(\ref{ppmthree}) and
(\ref{afreetop}--\ref{aev}). Similar field theories 
have been proposed in the past to explain the supeconductivity of
high temperature superconductors without breaking the $P$ and $T$
invariance, see \cite{wilczek}. The analogy between directed polymers and 
vortex lines 
has
been studied in connections with high $T_C$ superconductors in
\cite{nelson}. 
As in  superconductors of type II, also in the present case 
attractive and repulsive forces appear, which vanish at some self-dual point
of the theory. What is remarkable here, is that these interactions 
do not need the introduction of any potential and
are purely related to the topological constraints imposed on the
trajectories of the original polymer system.
Indeed, they
remain even if the short-range interactions are switched off as shown
in Eq.~(\ref{selfdual}). 
From the condition (\ref{selfdualcond}), which determines the
existence of the self-dual point or not, it is possible to predict
that $2s-$plats consisting of homogeneous polymers should have a
profoundly different behavior than their counterparts built out of
block copolymers. As a matter of fact, 
we have seen that the  physical characteristics
of the $2s$
directed polymers 
into which the $2s-$plat has been split
are described by the constants
$g_{a,I_a}$. In particular, the rigidity of the trajectories
may be specified by choosing the $g_{a,I_a}'$s appropriately.
Clearly, from Eq.~(\ref{selfdualcond}) it turns out that the self-dual
point is attained only if these constants are all equal, implying that
either the polymer rings are homopolymers or their subtrajectories
$\Gamma_{a,I_a}$ contain monomers of different types but, after
averaging over the distance of many monomer sizes, they have
identical physical properties. 
We have also derived the equations of motion
that minimize the action the anyon field theory in the case of a
$4-$plat. These equation describe the self-dual point of the
two-component anyon gas. 
After many simplifications, the relevant degrees of freedom can be
derived by solving the sinh-Gordon equation (\ref{sinhgordon}) and the
Laplace equations (\ref{fdsfsd1}--\ref{fdsfsd2}).

From the polymer point of view, the physical
meaning of 
the self-duality is unclear, because here only static conformations
have been considered. In principle, these static solutions could
become physically relevant in the case
of a very long $4-$plat in which
the monomer concentration does not depend  on
the height $z$. What is however more important, is that
Eq.~(\ref{sinhgordon}), which determines the static density of
monomers of type 1, is a sinh-Gordon equation identical to that
obtained in \cite{dunajski} for the static vortices of a relativistic
abelian Higgs model on a special type of Riemann surfaces.  
This analogy between field theories on Riemann surfaces and polymers,
together with the connections between linked polymers and multicomponent anyon
systems, that are related both to topological quantum computing and to
high-$T_C$ superconductors, should be further explored. It is true that
for topological computing nonabelian anyon systems are necessary, while
our discussion has been limited to the abelian case. However, this
limitation is only apparent. In principle, instead of the Gauss
linking invariant, we could have used much more refined topological
invariants that would have led to nonabelian anyon field theories
\cite{kleinert,vilgiskholodenko}. Up
to now, however, nobody has succeeded to formulate completely a
nonabelian theory of topological entanglement for polymer systems
based on such topological invariants, apart from a few exceptions
\cite{FFNOVA,FFJMP2003}.
Also the possibility of studying the statistical mechanics of knots
constructed from tangles should be investigated, because up to now
there is no analytical model which is able to describe the statistical
properties of knots. 
\section{Acknowledgments}
F. Ferrari would like to thank E. Szuszkiewicz for pointing out
Ref.~\cite{collins} and inspiring the present work.
We wish to thank
heartily also M. Pyrka, V. G. Rostiashvili and T. A. Vilgis for fruitful
discussions. 
The support of the Polish National Center of Science,
scientific project No. N~N202~326240, is gratefully acknowledged.
The simulations reported in this work were performed in part using the HPC
cluster HAL9000 of the Computing Centre of the Faculty of Mathematics
and Physics at the University of Szczecin. 
\begin{appendix}
\section{The length $L$ of a directed polymer as a function of the height}
In this Appendix we consider the partition function
\begin{equation}
{\cal Z}=\int\!{\cal D}\mathbf r(z)e^{-S}
\end{equation}
where $S$ is the action of the free open
polymer, whose trajectory $\Gamma$
is parametrized by means of the height $z$ defined 
in some interval $[\tau_0,\tau_1]$:
\begin{equation}
S=g
\int_{\tau_0}^{\tau_1}dz\left|\frac{d\mathbf r}{dz}\right|^2
\end{equation}
We want now to determine how the total length  of the curve $\Gamma$
depends on the constant parameter $g$. 
To understand what we mean by that, let us consider the standard
case of an ideal chain whose trajectory is
parametrized with the help of
the arc-length $\sigma$.
We denote with $a$ the average statistical length (Kuhn length)
of the $N$ segments  composing the polymer.
In the limit of large $N$ and small $a$ such that the product
$Na$ is constant, the total
length $L$ of the polymer 
satisfies the relation
\begin{equation}
L=Na \label{standlenght}
\end{equation}
We wish to obtain a similar identity connecting $L$ with $N$ and $g$
in the present situation, which is
 somewhat different.
To this purpose, we first dicretize the interval of integration
$[\tau_0,\tau_1]$ splitting it into $N$ small segments of length: 
\begin{equation}
\Delta z=\frac{\tau_1-\tau_0}{N}
\end{equation}
As a consequence, we may approximate the action as follows:
\begin{equation}
S\sim g\sum_{w=1}^{N}\left|\frac{\Delta \mathbf r_{w}}{\Delta
  z}\right|^{2}\Delta z
\end{equation}
where the symbol $\Delta \mathbf r_{w}$ means
\begin{equation}
\delta \mathbf r_{w}=\mathbf r_{w+1}-\mathbf r_{w}
\end{equation}
and
\begin{equation}
\mathbf r_{w}=\mathbf r(\tau_0+w\Delta z)
\end{equation}
The discretized partition function becomes thus the partition function
of a random chain composed by $N$ segments: 
\begin{equation}
{\cal Z}_{disc}=\int\!\prod_{w=1}^{N}d\mathbf
r_{w}e^{-\sum\limits_{w=1}^{N}g\frac{|\Delta \mathbf
    r_{w}|^{2}}{\Delta z}}  \label{zdiscr} 
\end{equation}
Using simple trigonometric arguments
it is easy to realize that
the  length of each segment is:
\begin{equation}
\Delta L=\sqrt{|\Delta \mathbf r_{w}|^{2}+\Delta z^{2}}
\end{equation}
This is of course an average length,
dictated by the fact that, from Eq.~\ceq{zdiscr},
the values of $|\Delta \mathbf{r}_{w}|$ should be gaussianly
distributed around the point: 
\begin{equation}
|\Delta \mathbf{r}_{w}|^{2}=\frac{\Delta z}{g}
\end{equation}
In the limit $\Delta z \rightarrow 0$,
 the distribution of length of $\Delta \mathbf{r}_{w}$ becomes the Dirac
 $\delta$-function: 
\begin{equation}
\lim_{\Delta z\rightarrow 0}\frac{1}{2}\sqrt{\frac{g}{\Delta
    z}}e^{-g|\Delta \mathbf{r}_{w}|^{2}/\Delta z}\sim \delta 
 \left(|\Delta \mathbf{r}_{w}|-\sqrt \frac{\Delta z}{g}\right)
\end{equation}
If $N$ is large enough, we can therefore conclude that the total
length of the chain $\Gamma$ is: 
\begin{equation}
L\sim N\Delta L=N\sqrt{\frac{\Delta z}{g}+\Delta z^{2}}
\end{equation}
Since $N\Delta z=\tau_1-\tau_0$, we get:
\begin{equation}
L^{2}=|\tau_1-\tau_0|^{2}+\frac{N(\tau_1-\tau_0)}{g} \label{pollength} 
\end{equation}
In the limit $N\rightarrow \infty$, while keeping the ratio
$\frac{N}{g}$ finite, Eq.~(\ref{pollength}) 
 becomes the desired relation between the length of $\Gamma$ and
$g$ which replaces Eq.~\ceq{standlenght}. 

\section{The expression of the Gauss linking invariant in the Coulomb gauge.}

\begin{figure}[bpht]
\centering
\includegraphics[scale=.5]{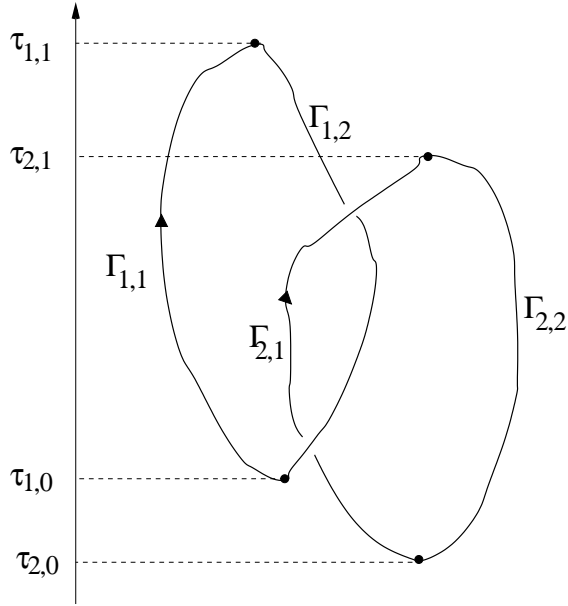}\\
\caption{Example of configuration of a $4-$plat.}\label{fig5}
\end{figure}

To fix the ideas, we will study here the particular case of a $4-$plat.
In the partition function \ceq{ppm} we isolate only the terms in which
the BF fields appear, 
because the other contributions are not connected to topological
constraints and thus are not relevant. 
As a consequence, we have just to compute the following partition function:
\begin{equation}
{\cal Z}_{CS,CG}(\lambda)=\int\!{\cal D}B_{\mu} {\cal
  D}C_{\mu}e^{-iS_{BF,CG}-S_{tot}} \label{ZCSaux} 
\end{equation}
where the BF action in the Coulomb gauge $S_{BF,CG}$ has
been already defined in Eq.~\ceq{CSCG} 
and $S_{top}$ has been given in Eq.~\ceq{stop}.
In the case of a $4-$plat, $S_{top}$ becomes:
\begin{eqnarray}
S_{top}&=&i\lambda
\int_{\tau_{1,0}}^{\tau_{1,1}}dt\left[\frac{dx_{1,1}^{\mu}(t)}{dt}
B_{\mu}(\mathbf{r}_{1,1}(t),t)- 
\frac{dx_{1,2}^{\mu}(t)}{dt}B_{\mu}(\mathbf{r}_{1,2}(t),t) \right]\nonumber \\
&+&\frac{i\kappa}{2\pi}\int_{\tau_{2,0}}^{\tau_{2,1}}dt
\left[\frac{dx_{2,1}^{\mu}(t)}{dt}C_{\mu}(\mathbf{r}_{2,1}(t),t) 
-\frac{dx_{2,2}^{\mu}(t)}{dt}C_{\mu}(\mathbf{r}_{2,2}(t),t)\right]
\end{eqnarray}
where we recall that $x_{a,I}^{\mu}(t)=(\mathbf{r}_{a,I}(t),t)$,
$a=1,2$, $I=1,2$. 
Using the Chern-Simons propagator of Eqs.~\ceq{propone}-\ceq{proptwo},
it is easy to evaluate the path integral over 
the gauge fields in Eq.~\ceq{ZCSaux}.
The result, after two simple Gaussian integrations, is:
\begin{equation}
Z_{BF,CG}(\lambda)=\exp{\left\lbrace
  \frac{i\lambda}{2\pi}\sum_{I,J=1}^2(-1)^{I+J-2}
\varepsilon_{ij} 
\int_{\tau_0}^{\tau_1}d(x_{1,I}^{i}(t)-x_{2,J}^{i}(t))
\frac{(x_{1,I}^{j}(t)-x_{2,J}^{j}(t)) 
}{\left|\mathbf{r}_{1,I}(t)-\mathbf{r}_{2,J}(t)
\right|^{2}}\right\rbrace} \label{resZCS} 
\end{equation}
In the above equation we have put for simplicity:
\begin{eqnarray}
\tau_0&=&\max[\tau_{1,0},\tau_{2,0}]
\nonumber\\
\tau_{1}&=&\min[\tau_{1,1},\tau_{2,1}]
\end{eqnarray}
For instance, if the polymer configurations are as in Fig.~\ref{fig5}, we
have that $\tau_0=\tau_{1,0}$ and $\tau_1=\tau_{2,1}$.
Moreover, we remember that in our notation 
$\mathbf{r}_{a,I}(t)=(x_{a,I}^{1}(t),x_{a,I}^{2}(t))$.
Apparently, the elements 
of the trajectories $\Gamma_{1}$ and $\Gamma_{2}$ which lie below
$\tau_0$ and above $\tau_1$ 
do not take the part in the topological interactions. Thus is due to
the presence of the Dirac $\delta$-function $\delta(t-t^{\prime})$ 
in the components of the Chern-Simons propagator \ceq{propone}-\ceq{proptwo}.
However, we will see later that also the contributions
of these missing parts are present in 
the expression of $Z_{CS}(\lambda)$.
In order to proceed, we notice that the exponent of the right hand
side of Eq.~\ceq{resZCS} consists 
 in a sum of integrals over the time $t$ of the kind:
\begin{equation}
D_{1,I;2,J}(\tau_1)-D_{1,I;2,J}(\tau_0)=
\varepsilon_{ij}\int_{\tau_0}^{\tau_1}d\left(x_{1,I}^{i}(t)-x_{2,J}^{i}(t)
\right)\frac{(x_{1,I}^{j}(t)-x_{2,J}^{j}(t))
}{\left|\mathbf{r}_{1,I}(t)-\mathbf{r}_{2,J}(t)\right|^{2}}
\end{equation}
The above integrals can be computed exactly. It is in fact
well known that the function $D_{1,I;2,J}(t)$ is the winding angle of the
vector $\mathbf r_{1,I}(t)-\mathbf r_{2,J}(t)$ at time $t$: 
\begin{equation}
D_{1,I;2,J}(t)=\arctan \left(\frac{x_{1,I}^{1}(t)-x_{2,J}^{1}(t)}{x_{1,I}^{2}(t)-x_{2,J}^{2}(t)}\right)\label{dfunction}
\end{equation}
Thus, the quantity $D_{1,I;2,J}(\tau_1)-D_{1,I;2,J}(\tau_0)$ is a
difference of winding angles which measures  
how many times the trajectory $\Gamma_{1,I}$ turns around the trajectory $\Gamma_{2,J}$ in the slice of time $\tau_0\leq t\leq \tau_1$.
At this point, without any loss of generality,
we suppose that the configurations of the curves
$\Gamma_1$ and $\Gamma_2$ 
is such that the maxima and minima $\tau_{a,I}$ are ordered as follows:
\begin{equation}
\tau_{2,0}<\tau_{1,0}<\tau_{2,1}<\tau_{1,1}
\end{equation}
As example of loop configurations that respect this ordering is
given in Fig.~\ref{fig5}. 
As a consequence, we have:
\begin{equation}
\tau_0=\tau_{1,0} \qquad \mbox{and}\qquad \tau_1=\tau_{2,1}
\end{equation}
Now we notice that the logarithm of the gauge partition function ${\cal
  Z}_{BF,CG}(\lambda)$ in Eq.~\ceq{resZCS} contains a sum of
differences of the winding angles defined in Eq.~(\ref{dfunction}): 
\begin{eqnarray}
\frac{2\pi\log{{\cal
      Z}_{BF,CG}}(\lambda)}
{i\lambda}&=&\left[D_{1,1;2,1}(\tau_{2,1})-D_{1,1;2,1}(\tau_{1,0})+ 
D_{1,2;2,2}(\tau_{2,1})-D_{1,1;2,2}(\tau_{2,1})\right.\nonumber \\
&+&\left.D_{1,2;2,1}(\tau_{1,0})-D_{1,2;2,1}(\tau_{2,1})+
D_{1,1;2,2}(\tau_{1,0})-D_{1,2;2,2}(\tau_{1,0})\right] \label{logZCS}
\end{eqnarray}
Further, assuming that the curves $\Gamma_1$ and $\Gamma_2$ are oriented as in
Fig.~\ref{fig5}. 
if we start from the minimum point at $\tau_0=\tau_{1,0}$, we can
 isolate in the right hand side of Eq.~\ceq{logZCS} 
the following four contributions:
\begin{enumerate}
\item
In the time slice $\tau_{1,0}\leq t\leq \tau_{2,1}$
the angle which measures the winding of the trajectory $\Gamma_{1,1}$
around the trajectory $\Gamma_{2,1}$ is given by the difference
$D_{1,1;2,1}(\tau_{2,1})-D_{1,1;2,1}(\tau_{1,0})$. 
\item
In the region $\tau_{2,1}\leq t\leq \tau_{1,1}$ only the trajectory
$\Gamma_1$ continues to evolve, 
going first upwards with the subtrajectory $\Gamma_{1,1}$ and then downwards
with $\Gamma_{1,2}$.
After this evolution, the winding angle 
between the two trajectories $\Gamma_1$ and $\Gamma_2$ has changed by
the quantity $D_{1,2;2,2}(\tau_{2,1})-D_{1,1;2,2}(\tau_{2,1})$.
\item
Next, in the region $\tau_{2,1}\geq t\geq \tau_{1,0}$, the winding
angle which measures how many 
times the subtrajectory $\Gamma_{1,2}$ winds up around 
$\Gamma_{2,2}$ is given by the 
difference $D_{1,2;2,1}(\tau_{1,0})-D_{1,2;2,1}(\tau_{2,1})$.
\item
Finally, in the region $\tau_{1,0}\geq t\geq \tau_{2,0}$ only the
second trajectory $\Gamma_2$ continues 
to evolve, going first downwards with the curve $\Gamma_{2,2}$ and
then upwards with $\Gamma_{2,1}$. 
The net effect of this evolution is that the 
winding angle between
$\Gamma_1$ and $\Gamma_2$ changes
by the quantity $D_{1,1;2,2}(\tau_{1,0})-D_{1,2;2,2}(\tau_{1,0})$.
\end{enumerate}
It is thus clear that the right hand side of Eq.~\ceq{logZCS}, apart
from a proportionality factor $i\lambda$,
counts how many times the trajectory $\Gamma_{1}$ winds around the
trajectory $\Gamma_{2}$. If we wish to identify the quantity in the
right hand side of Eq.~\ceq{logZCS} with the Gauss linking number
$\chi(\Gamma_1,\Gamma_2)$, we should check
for consistency that it  takes only integer values as
the Gauss linking number does. 
Indeed, it is easy to see that, modulo $2\pi$, the
following identities are holding: 
\begin{eqnarray}
D_{1,1;2,1}(\tau_{2,1})&=&D_{1,1;2,2}(\tau_{2,1})\nonumber\\
D_{1,1;2,2}(\tau_{1,0})&=&D_{1,2;2,2}(\tau_{1,0}) \nonumber \\
D_{1,2;2,2}(\tau_{2,1})&=&D_{1,2;2,1}(\tau_{2,1})\nonumber\\
D_{1,1;2,1}(\tau_{1,0})&=&D_{1,2;2,1}(\tau_{1,0})
\end{eqnarray}
For example, the first of the above equalities states that the 
angle formed by the vector $\mathbf r_{1,1}-\mathbf r_{2,1}$
connecting the subtrajectories $\Gamma_{1,1}$ and $\Gamma_{2,1}$ at
the height $\tau_{2,1}$ is equal to the angle
formed by the vector $\mathbf r_{1,1}-\mathbf r_{2,2}$
connecting the subtrajectories $\Gamma_{1,1}$ and $\Gamma_{2,2}$ at
the same height. The reason of this identity is trivial: At that
height, the subtrajectories $\Gamma_{2,1}$ and $\Gamma_{2,2}$ are
connected together at the same point.
Applying the above relations to Eq.~\ceq{logZCS}, one may prove that:
\begin{equation}
\frac{2\pi\log{{\cal Z}_{BF,CG}}(\lambda)}{i\lambda}=0\qquad\qquad\mod 2\pi
\end{equation}
As a consequence, we can write:
\begin{equation}
{\cal Z}_{BF,CG}(\lambda)=e^{i\lambda\chi(\Gamma_1,\Gamma_2)} \label{Z}
\end{equation}
where $\chi(\Gamma_1,\Gamma_2)$ is the Gauss linking number.
Concluding, the above analysis shows that also in the Coulomb gauge the
BF fields in the polymer partition function
\ceq{ppm} fix the topological constraints \ceq{topconst} correctly, in
full consistency
with the results obtained in the covariant gauges. 
Of course this consistency was expected due to gauge invariance.
Yet, it is interesting that, using the Coulomb gauge, one may express
the Gauss 
linking number invariant in a way that is quite different from the usual
form given in Eq.~\ceq{ln}.
\section{From polymers to anyon field theories}
In this Appendix the passage from
the polymer partition function
\ceq{ppm} to the field theoretical formulation of Eq.~(\ref{ppmthree})
is performed. To this purpose, we have to
integrate over all polymer trajectories $\mathbf
r_{a,I_a}(t_{a,I_a})$, 
$a=1,2$ and $I_a=1,\ldots,2s_{a}$. 
The standard procedure to pass to field theory in polymer physics
consists in introducing
auxiliary
fields. This
procedure works of course also in the present case, but it is
complicated by the splitting of the trajectories $\Gamma_{a}$ into $2s_{a}$
subtrajectories.
First of all, we have to introduce external sources for each
subtrajectory as follows:
\begin{equation}
J_{a,I_a}(\mathbf x,t)=\int_{\tau_{a,I_a-1}}^{
\tau_{a,I_a}}dt_{a,I_a}\delta(\mathbf x-\mathbf
r_{a,I_a}(t_{a,I_a}))\delta(t-t_{a,I_a})(-1)^{I_a-1} 
\end{equation}
Here the coordinates
$(\mathbf x,t)$ are allowed to span the whole $\mathbb{R}^3$ space.
Now it is possible to write the following identity:
\begin{eqnarray}
&&\exp\left[ -\int_{\tau_{a,I_a-1}}^{\tau_{a,I_a}}dt_{a,I_a}
\int_{\tau_{b,J_b-1}}^{\tau_{b,J_b}}
dt_{b,J_b}(-1)^{I_a+J_b-2}V(\mathbf
 r_{a,I_a}(t_{a,I_a})-\mathbf r_{b,J_b}(t_{b,I_b}))\right]\nonumber
\\
&&
=\exp\left[ -\int \!d^{2}\mathbf xd^{2}\mathbf ydt d
  t'J_{a,I_a}(\mathbf x,t) 
V(\mathbf x,\mathbf y)\delta(t-t^{\prime})
J_{b,J_b}(\mathbf y,t^{\prime})\right]
\label{typeintegral}
\end{eqnarray}
where $a,b=1,2$, $I_a=1,\ldots,2s_{a}$, $J_b=1,\ldots,2s_{b}$.
Clearly, the right hand side of the above equation can be interpreted as
the generating functional of a free scalar field theory with propagator
$G(\mathbf x,\mathbf y;t,t^{\prime})=V(\mathbf x,\mathbf
y)\delta(t-t^{\prime})$.
At this point we notice that the weight $e^{-S_{EV}}$ that takes into
account
the
short-term interactions in the partition function \ceq{ppm} is a product of
exponents of the kind given in Eq.~\ceq{typeintegral}.
Thus, $e^{-S_{EV}}$ coincides formally with the generating functional
of a multi-component scalar field theory.
The minimum number of scalar fields that is necessary to express
$e^{-S_{EV}}$ as a generating
functional is
$2s_{a}+2s_{b}+2$. Let's call these fields
$\varphi_{1,I_a}(\mathbf x,t),\varphi_{2,J_b}(\mathbf x,t)$,
$I_a=1,\ldots,2s_{a}$, $J_b=1,\ldots,2s_{b}$ and 
$\phi_{1}(\mathbf x,t),\phi_{2}(\mathbf x,t)$.
$\phi_{1}(\mathbf x,t)$ and $\phi_{2}(\mathbf x,t)$ will be
responsible for the interaction between monomers belonging to
different loops, while the $\varphi_{1,I_a}(\mathbf
x,t)$'s and $\varphi_{2,J_b}(\mathbf x,t)$'s will take into
account the interactions of monomers belonging to the same loop.

Remembering that in the present case $V(\mathbf x, \mathbf
y)=V_0\delta(\mathbf x-\mathbf y)$, it is possible to verify
the validity of the following identity:
\begin{eqnarray}
e^{-S_{EV}}&=&
\int\!{\cal D}
\phi_{1}{\cal
  D}
\phi_{2}\exp\left[-\frac1{V_0} 
\int d^2\mathbf xdt\phi_{1}(\mathbf x,t)\phi_{2}(\mathbf
x,t)\right] \nonumber\\
&\times&
\prod_{I=1}^{2s_1}\int{\cal D}\varphi_{1,I}(\mathbf
 x,t)
\prod_{J=1}^{2s_2}\int{\cal D}\varphi_{2,J}(\mathbf
 x,t)\nonumber\\
&\times&\exp{\left[-\frac{1}{2V_{0}}\int d^{2}\mathbf
 xdt
\left(
\sum\limits^{2s_1}_{I_1,I_1'=1}\varphi_{1,I_1}(\mathbf
 x,t)\varphi_{1,I'_1}(\mathbf
 x,s)\alpha_{I_1 I_1'}^{-1}
\right)\right]}
\nonumber\\
&\times&\exp{\left[-\frac{1}{2V_{0}}\int d^{2}\mathbf
 xdt
\left(
\sum\limits^{2s_2}_{J_2,J_2'=1}\varphi_{2,J_2}(\mathbf
 x,t)\varphi_{2,J'_2}(\mathbf
 x,s)\alpha_{J_2 J_2'}^{-1}
\right)\right]}
\nonumber\\
 &\times&
 \exp\left[-i\sum\limits^{2s_1}_{I=1}\int d^{2}\mathbf
 xdt J_{1,I}(\mathbf x,t))(\phi_{2}(\mathbf
 x,t)+\varphi_{1,I}(\mathbf 
 x,t))\right] \nonumber
\\
&\times&
\exp\left[{-i\sum\limits^{2s_2}_{J=1}\int d^{2}\mathbf xdtJ_{2,J}(\mathbf
 x,t))(\phi_1(\mathbf x,t)+\varphi_{2,J}(\mathbf
 x,t))}\right]\label{usingequation} 
\end{eqnarray}
where $\alpha_{I_a,I_a'}$, $I_a,I_a'=1,\ldots,2s_a$, is the
off-diagonal matrix 
\begin{equation}
\alpha_{I_aI'_a}=\left\{
\begin{array}{ccc}
0&\mbox{if}&I_a=I_a'\\
1&\mbox{if}&I_a\ne I_a'
\end{array}
\right.
\end{equation}
and $\alpha_{I_aI'_a}^{-1}$ represents its inverse.
Using equation (\ref{usingequation}), the partition function \ceq{ppm}
becomes: 
\begin{eqnarray}
{\cal Z}(\lambda)&=&\int\!{\cal D}B_{\mu}{\cal D}C_{\mu}e^{-iS_{BF}}\int
 {\cal D}\phi_{1}{\cal D}\phi_{2}e^{-\frac{1}{V_{0}}\int
 d^{2}\mathbf xdt\phi_{1}(\mathbf x,t)\phi_{2}(\mathbf
 x,t)}\nonumber\\
&\times&\prod_{I=1}^{2s_1}{\cal D}\varphi_{1,I}(\mathbf x,t)
\prod_{J=1}^{2s_2}{\cal D}\varphi_{2,I}(\mathbf x,t)
\nonumber\\
&\times&
\exp{\left[-\frac{1}{2V_{0}}\int d^{2}\mathbf
 xdt
\left(
\sum\limits^{2s_1}_{I_1,I_1'=1}\varphi_{1,I_1}(\mathbf
 x,t)\varphi_{1,I'_1}(\mathbf
 x,s)\alpha_{I_1 I_1'}^{-1}
+\sum\limits^{2s_2}_{J_2,J_2'=1}\varphi_{2,J_2}(\mathbf
 x,t)\varphi_{2,J'_2}(\mathbf
 x,s)\alpha_{J_2 J_2'}^{-1}
\right)\right]}
\nonumber\\
&\times& \left[\prod_{I=1}^{2s_1-1}\int _{\mathbf
 r_{1,I}(\tau_{1,I-1})}^{\mathbf
 r_{1,I}(\tau_{1,I})}\!{\cal D}\mathbf
 r_{1,I}(t_{1,I})\right]
\int _{\mathbf
 r_{1,2s_1}(\tau_{1,2s_1-1})}^{\mathbf
 r_{1,2s_1}(\tau_{1,0})}{\cal D}\mathbf
 r_{1,2s_1}(t_{1,2s_1})
\nonumber\\
&\times& \left[\prod_{J=1}^{2s_2-1}\int _{\mathbf
 r_{2,I}(\tau_{2,J-1})}^{\mathbf
 r_{2,J}(\tau_{2,J})}\!{\cal D}\mathbf
 r_{2,J}(t_{2,J})\right]
\int _{\mathbf
 r_{2,2s_2}(\tau_{2,2s_2-1})}^{\mathbf
 r_{2,2s_2}(\tau_{2,0})}{\cal D}\mathbf
 r_{2,2s_2}(t_{2,2s_2})
e^{-S_{eff}} 
\label{ppmtwo}
\end{eqnarray}
with
\begin{eqnarray}
S_{eff}&=&
\sum_{I=1}^{2s_{1}}
\int_{\tau_{1},I-1}^{\tau_{1},I}dt_{1,I}
\left[(-1)^{I-1}
g_{1,I}\left|\frac{d\mathbf
 r_{1,I}}{dt_{1,I}}\right|^{2}+
(-1)^{I-1}i(\phi_{2}(\mathbf
 r_{1,I}(t_{1,I}),t_{1,I})+\varphi_{1,I}(\mathbf r_{1,I}(t_{1,I}),t_{1,I}))\right.
\nonumber\\
&+&\left.i\lambda B_3(\mathbf r_{1,I}(t_{1,I}),t_{1,I})+i\lambda
\frac{d\mathbf r_{1,I}}{dt_{1,I}}\cdot \mathbf 
B(\mathbf r_{1,I}(t_{1,I}),t_{1,I})\right]
\nonumber\\
&+&\sum_{J=1}^{2s_{2}}\int_{\tau_{2},J-1}^{\tau_{2},J}dt_{2,J}
\left[(-1)^{J-1}g_{2,J}\left|\frac{d\mathbf 
 r_{2,J}}{dt_{2,J}}\right|^{2}+(-1)^{J-1}i(\phi_{1}(\mathbf
 r_{2,J}(t_{2,J}),t_{2,J})+\varphi_{2,J}(\mathbf
 r_{2,J}(t_{2,J}),t_{2,J}))\right.
\nonumber\\
&+&\left.i\kappa C_3(\mathbf
 r_{2,J}(t_{2,J}),t_{2,J})+\frac{i\kappa}{2\pi} \frac{d\mathbf
   r_{2,J}}{dt_{2,J}}\cdot 
 \mathbf C(\mathbf r_{2,J}(t_{2,J}),t_{2,J})\right]
\end{eqnarray}
Of course, the ends of the trajectories $\Gamma_{1,I}$ and
$\Gamma_{2,J}$ appearing in the limits of path integration
over $\mathbf r_{1,I}(t_{1,I})$ and $\mathbf r_{2,J}(t_{2,J})$ in
Eq.~\ceq{ppmtwo} are not all independent, because they
 are subjected to the constraints \ceq{bcondtwoa}
and \ceq{bcondtwo}. We will get rid of these constraints later when
passing to the field theoretical representation.

Let us use at this point the so-called complex replica fields defined
in Eq.~(\ref{replicafielddef}).
Then, the path integrals over the trajectories $\mathbf r_{a,I}$ may be
rewritten as follows \cite{FFAnnPhys2002}:
\begin{eqnarray}
\!\!\!\!\!\!\!\!&&
 \left[\prod_{I=1}^{2s_1-1}\int _{\mathbf
 r_{1,I}(\tau_{1,I-1})}^{\mathbf
 r_{1,I}(\tau_{1,I})}\!{\cal D}\mathbf
 r_{1,I}(t_{1,I})\right]
\int _{\mathbf
 r_{1,2s_1}(\tau_{1,2s_1-1})}^{\mathbf
 r_{1,2s_1}(\tau_{1,0})}{\cal D}\mathbf
 r_{1,2s_1}(t_{1,2s_1})
\nonumber\\
&& \left[\prod_{J=1}^{2s_2-1}\int _{\mathbf
 r_{2,I}(\tau_{2,J-1})}^{\mathbf
 r_{2,J}(\tau_{2,J})}\!{\cal D}\mathbf
 r_{2,J}(t_{2,J})\right]
\int _{\mathbf
 r_{2,2s_2}(\tau_{2,2s_2-1})}^{\mathbf
 r_{2,2s_2}(\tau_{2,0})}{\cal D}\mathbf
 r_{2,2s_2}(t_{2,2s_2})
e^{-S_{eff}} 
= \nonumber\\
\!\!\!\!\!\!\!\!&&\lim_{n_1\rightarrow 0}
\left[\prod_{I=1}^{2s_1-1}
\int\!{\cal D}\vec{\Psi}_{1,I} {\cal D}\vec{\Psi}_{1,I}^{*}
\psi_{1,I}^1(\mathbf r_{1,I}(\tau_{1,I-1}),\tau_{1,I-1})
\psi_{1,I}^{1*}(\mathbf r_{1,I}(\tau_{1,I}),\tau_{1,I})
\right]
 \nonumber\\
&\times&
\psi^{1*}_{1,2s_1}(\mathbf r_{1,2s_1-1}(\tau_{1,2s_1-1}),\tau_{1,2s_1-1})
\psi^{1}_{1,2s_1}(\mathbf r_{1,1}(\tau_{1,0}),\tau_{1,0})
\nonumber\\
&\times&
\prod_{I=1}^{2s_1}
\exp\Bigg\lbrace-
\int_{\tau_{1},I-1}^{\tau_{1},I}\!dt(-1)^{I-1}
\int \!d^{2}\mathbf x\vec{\Psi}_{1,I}^{*}(\mathbf x,t)
\cdot\left[\frac{\partial}{\partial t}-\frac{1}{4g_{1,I}}
(\boldsymbol \nabla -i\lambda(-1)^{I-1}\mathbf B(\mathbf
  x,t))^2
\right.
\nonumber\\
&+&
\left.
i\lambda(-1)^{I-1}
B_3(\mathbf x,t)
+i(\phi_2(\mathbf x,t)+
\varphi_{1,I}
(\mathbf x,t))\phantom{\frac{\partial}{\partial}}\!\!\!\!\!\right]
\vec \Psi_{1,I}^{*}(\mathbf x,t)\Bigg\rbrace
\nonumber\\
&\times&
\lim_{n_{2}\rightarrow 0}
\prod_{J=1}^{2s_{2}-1}
\int\!{\cal D}\vec{\Psi}_{2,J} {\cal D}\vec{\Psi}_{2,J}^*
\psi_{2,J}^1(\mathbf r_{2,J-1}(\tau_{2,J-1}),\tau_{2,J-1})
\psi_{2,J}^{1*}(\mathbf r_{2,J}(\tau_{2,J}),\tau_{2,J})
\nonumber\\
&\times&
\psi^{1*}_{2,2s_2}(\mathbf r_{2,2s_2-1}(\tau_{2,2s_2-1}),\tau_{2,2s_2-1})
\psi^{1}_{2,2s_2}(\mathbf r_{2,1}(\tau_{2,0}),\tau_{2,0})
\nonumber\\
&\times&\prod_{J=2}^{2s_2}\exp\Bigg\lbrace-
\int_{\tau_{2},J}^{\tau_{2},J+1}\!dt(-1)^{J-1}
\int \!d^{2}\mathbf x\vec{\Psi}_{2,J}^{*}(\mathbf x,t)
\cdot\left[\frac{\partial}{\partial t}-\frac{1}{4g_{2,J}}
(\boldsymbol\nabla -\frac{i\kappa}{2\pi}(-1)^{J-1}\mathbf C(\mathbf x,t))^{2}
\right.
\nonumber\\
&+&\left.\frac{i\kappa}{2\pi}(-1)^{J-1}C_3(\mathbf x,t)
+i(\phi_{1}(\mathbf x,t)+\varphi_{2,I}(\mathbf x,t))\right]
\vec \Psi_{2,J}(\mathbf x,t)\Bigg\rbrace
\end{eqnarray}
Let us note that with the above choice of arguments of the fields
$\psi_{1,I}^{1*},\psi_{1,I}$ 
and $\psi_{2,J}^{1*},\psi_{2,J}$, $I=1,\ldots,2s_1$,
$J=1,\ldots,2s_2$, the constraints
\ceq{bcondtwoa}
and \ceq{bcondtwo} are already taken into account.
Inserting this result in Eq.~\ceq{ppmtwo} and integrating out the
auxiliary fields $\varphi_{1,I},\varphi_{2,J},\phi_{1}$ and $\phi_{2}$, we
obtain the final expression of the polymer partition function given by
Eqs.~(\ref{ppmthree}), (\ref{afreetop}) and (\ref{aev}).
\end{appendix}


\begin{thebibliography}{99}
\bibitem{grosberg} A. Yu. Grosberg Phys.-Usp. {\bf 40},  12  (1997).
\bibitem{dna} W. R. Taylor, Nature (London) {\bf 406}, 916 (2000).
\bibitem{katritch96} V. Katritch, J. Bednar, D. Michoud, R. G.  Scharein, 
J. Dubochet, A. Stasiak, Nature {\bf 384}, 142 (1996). 
\bibitem{katritch97} V. Katritch, W. K.  Olson, P. Pieranski, J. Dubochet 
and A. Stasiak, Nature {\bf 388}, 148 (1997). 
\bibitem{krasnow} M. A. Krasnow, A. Stasiak, S. J. Spengler, F. Dean, 
T. Koller and N. R. Cozzarelli, Nature {\bf 304}, 559 (1983).
\bibitem{laurie} B. Laurie, V. Katritch, J. Dubochet and A. Stasiak,
  Biophys. Jour. {\bf 74}, 2815 (1998).
\bibitem{cieplak} J. I. Su{\l}kowksa, P. Su{\l}kowksa, P. Szymczak and
M. Cieplak, Phys. Rev. Lett. {\bf 100}, 058106 (2008).
\bibitem{marko} J. F. Marko, {\it Phys. Rev.} {\bf E 79} (2009), 051905.
\bibitem{liu} Z. Liu, E. L. Zechiedrich, and H. S. Chan, 
  Biophys. J. {\bf 90}, 2344 (2006). 
\bibitem{wasserman} S. A. Wasserman and N. R. Cozzarelli,
Science {\bf 232}, 951 (1986).
\bibitem{sumners} D. W. Sumners,  “Knot theory and DNA,” in New Scientific Applications
of Geometry and Topology, edited by  
D.
W. Sumners, Proceedings of Symposia in Applied Mathematics, Vol. 45,
͑American Mathematical Society, Providence, RI, 1992, 39. 
\bibitem{vologodski} A. V. Vologodski, ̆ A. V. Lukashin, M. D. Frank-Kamenetski ̆ and V. V. Anshelevich, Zh.
Eksp. Teor. Fiz. {\bf 66}, 2153 (1974); Sov. Phys. JETP 39, 1059 (1975);
M. D. Frank-Kamenetskii, A. V. Lukashin and A. V. Vologodskii, Nature
(London) {\bf 258},
398 (1975).
\bibitem{orlandini} E. Orlandini, S. G. Whittington, Rev. Mod. Phys. {\bf 79},
611 (2007);
 C. Micheletti, D. Marenduzzo, and E. Orlandini, Phys. Reports {\bf
   504}, 1 (2011).
\bibitem{levene} S. D. Levene, C. Donahue, T. C. Boles and N. R. Cozzarelli,
Biophys. J. {\bf 69} (1995) 1036.
\bibitem{kurt} T. Vettorel, A. Yu. Grosberg and K. Kremer,
  Phys. Biol. {\bf 6}, 025013  (2009).
\bibitem{mehran} P. Virnau, Y. Kantor and M. Kardar, J. Am. Chem. Soc. {\bf 127} (43),  15102 (2005).
\bibitem{pp2} P. Piera\'nski, S. Przyby{\l} and A. Stasiak, EPJ E
{\bf 6} (2), 123  (2001). 
\bibitem{yan} J. Yan, M. O. Magnasco and J. F. Marko, {\it Nature} {\bf
  401} (1999), 932.
\bibitem{arsuaga} J. Arsuaga, M. Vazquez, S. Trigueros, D. W. Sumners
  and J. Roca, {\it PNAS} {\bf 99} (2002), 5373.
\bibitem{arsuaga2} J. Arsuaga, M. Vazquez, P. McGuirk, S. Trigueros,
  D. W. Sumners 
  and J. Roca, {\it PNAS} {\bf 102} (2005), 9165.
\bibitem{metzler} R. Metzler, A. Hanke, P. G. Dommersnes, Y. Kantor
  and M. Kardar,
Phys. Rev. Lett. {\bf 88}, 188101 (2002).
\bibitem{pieranski} P. Pieranski, S. Clausen, G. Helgesen and A. T. Skjeltorp, 
Phys. Rev. Lett. {\bf 77},
1620 (1996).
\bibitem{diao} Y. Diao, C. Ernst and E. J. Janse van Rensburg in Ideal
  Knots, A. Stasiak, V.
Katritch and L. H. Kauffman (Eds), (World Scientific, Singapore, 1998)
p.52.
\bibitem{sumners2} D. W. Sumners, {\it Notices of the Am. Math. Soc.}
  {\bf 42} (5) 
  (1995), 528.
\bibitem{faddeev} L. Faddeev and A. J. Niemi,
{\it Nature} {\bf 387} (1997), 58.
\bibitem{birman} J. S. Birman, {\it Braids, links, and mapping class
  groups}, (Princeton University Press 1974).
\bibitem{darcy} I. K. Darch and R. G. Scharein, {\it Bioinformatics},
  {\bf 22} (14) (2006), 1790.
\bibitem{FFPLA2004} F. Ferrari, {\it Phys. Lett. } {\bf A323}, (2004),
  351, cond-mat/0401104.
\bibitem{dassarma} Das Sarma, S., M. Freedman, and C. Nayak,  {\it Topologi-
cal quantum computation}, Phys. Today {\bf 59} ͑(7͒) (2006), 32.
\bibitem{nayak} C. Nayak, S. H. Simon, A. Stern, M. Freedman and S. Das
  Sarma, {\it Rev. mod. Phys.} {\bf 80} (2008), 1083.
\bibitem{wilczek} F. Wilczek, {\it New kinds of quantum statistics},
article published in {\it The Spin}, 
{\it Progress in Mathematical Physics} {\bf  55} (2009), 61. 
\bibitem{goldman} V. Goldman, J. Liu and A. Zaslavsky, {\it
  Phys. Rev.} {\bf B71} (2005), 153303; F. Camino, F.Zhou and
  V. Goldman, {\it Phys. Rev. Lett.} {\bf 98} (2007), 076805.
\bibitem{anyonsexp2013} G. Ben-Shach, C. R. Laumann, I. Neder,
  A. Yacoby, and B. I. Halperin, {\it Phys. Rev. Lett.} {\bf 110}
  (2013), 106805.
\bibitem{gurarie} V. Gurarie, L. Radzihovsky, and A. V. Andreev,
{\it Phys. Rev. Lett.} {\bf 94} (2005), 230403.
\bibitem{dolev} M. Dolev, M. Heiblum, V. Umansky, A. Stern, and
  D. Mahalu, Nature 452 
(2008), 829; I. Radu, J. Miller, C. Marcus, 
M. Kastner, L. Pfeiffer, and K. West, Science 320 (2008), 899.
\bibitem{thomblau} M. Blau and G. Thompson, {\it Annals Phys.}
{\bf 205} (1991), 130.

\bibitem{abeliananyons} D. F. Milne, N. V. Korolkova and P. van Loock,
  {\it Phys. Rev. A.} {\bf 85} (5) (2012), 052325.
\bibitem{membranes} 
See e.~g. G. Decher, E. Kuchinka, H. Ringsdorf, J. Venzmer,
D. Bitter-Suermann and C. Weisgerber,  {\it Angew. Makromol. Chem.}, 
{\bf 71} (1989), 166;
J. Simon, M. K\"uhner, H. Ringsdorf and E. Sackmann, {\it
  Chem. Phys. Lipids.}
{\bf 76}  (2) (1995), 241;
G. Blume and  G. Cevc, {\it  Biochim. Biophys. Acta} {\bf 1029}  (1990),
91; D. D. Lasic, F. J. Martin, A. Gabizon, S. K. Huang, D. Papahadjopoulos;
{\it Biochim.
Biophys. Acta} {\bf 1070} (1991), 187.
\bibitem{edwards}   S. F. Edwards, {\it Proc.~Phys.~ Soc.}
 {\bf 91} (1967), 513;  {\it Proc.~Phys.~ Soc.}
 {\bf 92} (1967), 9.
\bibitem{FELAPLB1998} F. Ferrari and I. Lazzizzera,
{\it Phys. Lett.} {\bf B444} (1998), 167.
\bibitem{FELAJPA1999} F. Ferrari and I. Lazzizzera, 
{\it Jour. Phys. A}: {\it Math. Gen.} {\bf 32} (1999), 1347,
hep-th/9803008.
\bibitem{blauthompson} D. Birmingham, M. Blau, M. Rakowski and
  G. Thompson, {\it Phys. Rep.} {\bf 209} (1991), 129.
\bibitem{Kardar} M. Kardar, {\it J. Appl. Phys.} {\bf 61} (1987), 3601;
M. Kardar, G. Parisi, Y.-C. Zhang, {\it Phys. Rev. Lett.}
{\bf 56} (1986), 889;
M. Kardar, Y.-C. Zhang, {\it Phys. Rev. Lett.}
{\bf 58} (1987), 2087.
\bibitem{kamien} R. D. Kamien, P. Le Doussal and D. Nelson, {\it
  Phys. Rev.} {\bf A 45} (1992), 8727.
\bibitem{degennes} P. G. de Gennes, {\it Phys. Lett.} {\bf A38} (1972), 339;
J. des Cloiseaux, {\it Phys. Rev} {\bf A10} (1974), 1665;
V. J. Emery, {\it Phys. Rev.} {\bf B11} (1975), 239.
\bibitem{wilczek2} F. Wilczek, {\it Phys. Rev. Lett.} {\bf 69} (1992),
  132.
\bibitem{thompsonblau} M. Blau and G. Thompson,
Ann. Phys. 205 (1991), 130.
\bibitem{froehlichking} J. Froehlich and C. King, {\it Comm. Math. Phys.}
{\bf 126} (1) (1989), 167.
\bibitem{FFNOVA} 
F. Ferrari, {\it Topological field theories with non-semisimple gauge
group of symmetry and engineering of topological invariants},
chapter published in {\it Trends in Field
Theory Research}, O. Kovras (Editor), Nova Science Publishers (2005),
 	ISBN:1-59454-123-X. See also the reprint of this article in
 {\it Current Topics in Quantum Field Theory Research},
O. Kovras (Editor), Nova Science Publishers (2006),
ISBN: 1-60021-283-2.
\bibitem{FFAnnPhys2002} F. Ferrari, {\it Annalen der Physik} (Leipzig)
  {\bf 11} (2002) 4, 
255--290.
\bibitem{dunne} G. Dunne, 	
{\it Self-Dual Chern-Simons Theories}, Lecture Notes in Physics, New
Series M: Monographs, Vol. {\bf 36}, (Springer Verlag, 1995).
\bibitem{ferrari} F. Ferrari  and
I. Lazzizzera, 
{\it Nucl. Phys.} {\bf B559} (3) (1999), 673.
\bibitem{vilgisotto} M. Otto and T. A. Vilgis, {\it Phys. Rev. Lett.} {\bf 80} 
(1998), 881.
\bibitem{otto}  M. Otto, {\it J. Phys. A: Math. Gen.} {\bf 34} (12) (2001),
2539.
\bibitem{Brereton} M. G. Brereton, {\it Jour. Mol. Struct.} (Theochem),
{\bf 336} (1995), 191.
\bibitem{nelson} D. R. Nelson and H. S. Seung, {\it Phys. Rev. } {\bf
  B39} (1989), 9153.
\bibitem{dunajski} M. Dunajski, {\it Abelian vortices from
  Sinh--Gordon and Tzitzeica equations}, {\it Phys. Lett}
  {\bf B710} (2012), 236, arXiv:1201.0105v2 [hep-th]. 
\bibitem{kleinert} H. Kleinert, {\em Path
    Integrals in Quantum Mechanics, Statistics, Polymer Physics, and Financial
    Markets}, (World Scientific Publishing, 3rd Ed., Singapore, 2003).
\bibitem{vilgiskholodenko} A.L. Kholodenko and
    T.A. Vilgis, {\it Phys. Rep.} {\bf 298} (1998), 251.
\bibitem{FFJMP2003} F. Ferrari, 
{\it Jour. Math. Phys.} {\bf 44} (1) (2003), 138, hep-th/0210100.
\bibitem{collins} G. P. Collins, {\it Scientific American} {\bf 294}
  (4) (2006), 56.
\end{thebibliography}
\end{document}